\documentclass[11pt]{article}

\usepackage{graphicx,ifthen,color,algorithm,palatino}
\usepackage{amsmath,amsthm,amssymb,amsfonts,verbatim}
\usepackage{mathrsfs,accents,setspace}
\usepackage{subfigure,caption}
\usepackage{multirow,makecell}
\usepackage[authoryear]{natbib}
\newboolean{ShowFigures}
\newboolean{UnBlinded}
\newboolean{DoubleSpaced}
\setboolean{ShowFigures}{true}
\setboolean{UnBlinded}{true}
\setboolean{DoubleSpaced}{true}
\def\by{\boldsymbol{y}}

\def\bI{\boldsymbol{I}}

\def\bM{\boldsymbol{M}}

\def\bvarepsilon{\boldsymbol{\varepsilon}}

\def\bdeta{\boldsymbol{\eta}}
\def\btheta{\boldsymbol{\theta}}

\def\blambda{\boldsymbol{\lambda}}
\def\bmu{\boldsymbol{\mu}}
\def\bnu{\boldsymbol{\nu}}

\def\bpsi{\boldsymbol{\psi}}

\def\bTheta{\boldsymbol{\Theta}}
\def\bLambda{\boldsymbol{\Lambda}}

\def\bSigma{\boldsymbol{\Sigma}}

\def\diag{\mbox{diag}}

\def\Gfull{G_{\tiny\mbox{full}}}
\def\bzero{\boldsymbol{0}}

\def\simind{\stackrel{{\tiny \mbox{ind.}}}{\sim}}

\begin{document}

\ifthenelse{\boolean{DoubleSpaced}}{\setstretch{1.5}}{}

\centerline{\Large\bf Fitting Structural Equation Models via Variational Approximations}
\vskip7mm
\ifthenelse{\boolean{UnBlinded}}{
\centerline{\normalsize\sc By Khue-Dung Dang$\null^{1}$\footnotemark and Luca Maestrini$\null^{2,3}$\footnotemark[\value{footnote}]}
\footnotetext{The authors equally contributed to this work.}
\vskip5mm
\centerline{\textit{$\null^1$The University of Melbourne, $\null^2$University of Technology Sydney and}}
\vskip1mm
\centerline{\textit{$\null^3$Australian Research Council Centre of Excellence for Mathematical and Statistical Frontiers}}}{}
\vskip6mm
\centerline{\today}

\vskip6mm

\centerline{\large\bf Abstract}
\vskip2mm

{Structural equation models are commonly used to capture the relationship between sets of observed and unobservable variables. Traditionally these models are fitted using frequentist approaches but recently researchers and practitioners have developed increasing interest in Bayesian inference. In Bayesian settings, inference for these models is typically performed via Markov chain Monte Carlo methods, which may be computationally intensive for models with a large number of manifest variables or complex structures. Variational approximations can be a fast alternative; however, they have not been adequately explored for this class of models. We develop a mean field variational Bayes approach for fitting elemental structural equation models and demonstrate how bootstrap can considerably improve the variational approximation quality. We show that this variational approximation method can provide reliable inference while being significantly faster than Markov chain Monte Carlo.}


\section{Introduction}

Structural equation models are commonly used in social and behavioral sciences to study the structural relationship between manifest or observed variables, such as test scores or answers from a questionnaire, and latent constructs or unobservable variables \citep{kaplan2008structural}. The variation and covariation between the observed outcomes are explained by assuming that the observed variables are linked to the hidden factors. The models can then be used to test whether the data fit a hypothesized measurement model based on theory or previous research and check which observed variables are good indicators of the latent variables.
 
A basic structural equation model (SEM) usually consists of two components: a confirmatory factor analysis model that relates the latent variables to their corresponding manifest variables, taking measurement errors into account; a second component that regresses the endogenous and exogenous latent variables \citep{lee2007structural}. 
In this work we focus on the class of confirmatory factor analysis models.

Maximum likelihood and weighted least squares procedures for fitting SEMs are available in sofware packages such as \textsf{Mplus}, \textsf{OpenMX} and the package \textsf{lavaan} \citep{rosseel2012lavaan} from the R computing environment \citep{R2021}. These models are usually designed in such a way that the covariance matrix of the observed data contains all the model parameters, hence classical methods for SEMs typically analyze the sample covariance matrix  \citep{lee2007structural}. However, frequentist approaches may suffer from computational and theoretical problems for small sample sizes or non-normal data. Analyzing the maximum likelihood estimates of latent factor scores may be challenging as their asymptotic properties rely on the sample covariance matrix being asymptotically normal \citep{boomsma1982robustness,chou1991scaled}.
Recently, Bayesian fitting procedures have received more attention, as they overcome the issues of frequentist approaches when the number of observations is small and facilitate the adoption of more flexible model structures, such as those with cross-loadings and non-normal errors \citep{lee2007structural}. Bayesian approaches allow for inference on the latent variable factor scores through their posterior distribution and to incorporate useful prior information. For a complete discussion on the differences between Bayesian and frequentist approaches for fitting SEMs we refert to Chapter 4 of \cite{lee2007structural}.

Markov chain Monte Carlo (MCMC) methods have been largely studied and employed in mainstream research on SEMs. Recent versions of \textsf{Mplus} also support MCMC and the Bayesian version of the \textsf{lavaan} package, named \textsf{blavaan} \citep{merkle2018blavaan,merkle2021blavaan}, facilitates Bayesian statistical analysis via SEMs. However, MCMC methods may suffer from slow convergence and long running times, compared to frequentist approaches.
Variational approximations can be a fast alternative; however they have not been adequately explored for this class of models and the current work is limited to a variational method for a special case of latent curve models \citep{tiwari2016variational}. In this paper, we propose and study a mean field variational Bayes (MFVB) approach to fit a general class of SEMs. We also examine the use of bootstrap to improve the performance of MFVB. 

The paper is organized as follows. In Section \ref{sec:SEMbase} we introduce a simple SEM that is used to develop the MFVB scheme proposed in Section \ref{sec:mfvbSEM}. Section \ref{sec:realDataNoBoot} demonstrates how this variational method performs on a classic real data example. In Section \ref{sec:bootrapMethod} we describe a strategy to improve the accuracy of MFVB based on nonparametric bootstrap and in Section \ref{sec:simStudy} we show how bootstrap can substantially improve the accuracy of MFVB using a simulation study. Section \ref{sec:extension} shows that the basic MFVB algorithm can be easily extended to fit models with multiple factors and apply MFVB to a real data example. Section \ref{sec:conclusion} contains a discussion and remarks.

\section{A Simple Structural Equation Model}\label{sec:SEMbase}

Confirmatory factor analysis, and SEMs in general, are commonly used to test hypotheses about the relationship between multiple correlated responses and one or more latent factors. 
The relationship between a vector of observed variables $\by$ and latent variables $\bdeta$ is typically modeled as
\begin{equation*}
    \by=\bLambda\bdeta+\bvarepsilon,
\end{equation*}
where $\bLambda$ is a matrix of factor loadings and $\bvarepsilon$ is an error vector. For instance, $\by$ can be the outcome of a series of intelligence tests and psychologists may use the model to assess the hypothesis that these tests are related to a person's cognitive abilities, represented by $\bdeta$, which cannot be measured directly.

Consider the following structural equation model with a single factor:
\begin{equation}
    \begin{array}{c}
        \by_i\,\vert\,\bnu,\blambda,\eta_i,\bpsi\simind N\big(\bnu+\blambda \eta_i,\diag(\bpsi)\big),\,\,\, \eta_i\,\vert\,\sigma^2\simind N(0,\sigma^2),\,\,\, i=1,\ldots,n,\\[1ex]
        \lambda_j\,\vert\,\psi_j\simind N(\mu_{\lambda},\sigma^2_\lambda\psi_j),\,\,\,\nu_j\simind N(0,\sigma^2_\nu),\,\,\,\psi_j\simind\mbox{Inverse-}\chi^2(\kappa_{\psi},\delta_{\psi}),\,\,\, j=1,\ldots,m,\\[1.2ex]     \sigma^2\sim\mbox{Inverse-}\chi^2(\kappa_{\sigma^2},\delta_{\sigma^2}),
    \end{array}
\label{eq:simpleSEM}
\end{equation}
where $\by_i$ is a vector of $m$ observed outcomes for individual $i$ from a group of $n$ individuals; $\bnu\equiv[\nu_1,\ldots,\nu_m]$ and $\blambda\equiv[\lambda_1,\ldots,\lambda_m]$ are $m\times 1$ vectors of intercepts and loadings, or scores, for the latent factor $\eta_i$. The Normal priors on the elements of $\bnu$ and $\blambda$ have hyperparameters $\mu_{\lambda}$ and $\sigma_\nu,\sigma_\lambda>0$. For model identification purposes, the first element of $\blambda$ can be set to 1. Inverse-$\chi^2$ priors are assigned to the entries of $\bpsi\equiv[\psi_1,\ldots,\psi_m]$ and $\sigma^2$, using hyperparameters $\kappa_{\psi},\delta_{\psi},\kappa_{\sigma^2},\delta_{\sigma^2}>0$. 
Here a random variable $x$ has an Inverse-$\chi^2(\kappa,\delta)$ distribution with shape parameter $\kappa>0$
and scale parameter $\delta>0$ if its probability density function is $$p(x) = \{(\delta/2)^{\kappa/2}/\Gamma(\kappa/2)\}x^{-(\kappa+2)/2} \exp \{-\delta/(2x)\}I(x>0).$$
As elucidated in \cite{maestrini2021inverse}, the Inverse-$\chi^2$ priors on the variance parameters of model \eqref{eq:simpleSEM} may be replaced by hierarchical prior specifications involving distributions from the Inverse G-Wishart family that facilitate the imposition of arbitrarily non-informative priors on standard deviation parameters and guarantee close form updates in variational approximation algorithms.

Let $\bdeta\equiv[\eta_1,\ldots,\eta_n]$. The likelihood function arising from model \eqref{eq:simpleSEM} is
\begin{align*}
    &\hspace{-1cm}p(\by;\bnu,\blambda,\bdeta,\bpsi,\sigma^2)=\\
    &\prod_{i=1}^n\left[ (2\pi)^{-m/2}\prod_{j=1}^m\psi_j^{-1/2}\exp\left\{-\frac{1}{2}(\by_i-\bnu-\blambda\eta_i)^T\diag(1/\bpsi)(\by_i-\bnu-\blambda\eta_i)\right\}\right]\\
    \times&\prod_{j=1}^m\left\{(2\pi\sigma^2_\nu)^{-1/2}\exp\left(-\frac{\nu_j^2}{2\sigma^2_\nu}\right)\right\}\prod_{j=1}^m\left[(2\pi\sigma^2_\lambda\psi_j)^{-1/2}\exp\left\{-\frac{(\lambda_j-\mu_\lambda)^2}{2\sigma^2_\lambda\psi_j}\right\}\right]\\
    \times&\prod_{i=1}^n\left\{(2\pi\sigma^2)^{-1/2}\exp\left(-\frac{\eta_i^2}{2\sigma^2}\right)\right\}\prod_{j=1}^m\left[\left(\frac{\delta_\psi}{2}\right)^{\kappa_\psi/2}\left\{\Gamma\left(\frac{\kappa_\psi}{2}\right)\right\}^{-1}\psi_j^{-(\kappa_\psi+2)/2}\exp\left(-\frac{\delta_\psi}{2\psi_j}\right)\right]\\
    \times&\left(\frac{\delta_{\sigma^2}}{2}\right)^{\kappa_{\sigma^2}/2}\left\{\Gamma\left(\frac{\kappa_{\sigma^2}}{2}\right)\right\}^{-1}(\sigma^2)^{-(\kappa_{\sigma^2}+2)/2}\exp\left(-\frac{\delta_{\sigma^2}}{2\sigma^2}\right).
\end{align*}
Consequently, the log-likelihood function is
\begin{align*}
    &\log p(\by;\bnu,\blambda,\bdeta,\bpsi,\sigma^2)=-\frac{1}{2}\sum_{i=1}^n(\by_i-\bnu-\blambda\eta_i)^T\diag(1/\bpsi)(\by_i-\bnu-\blambda\eta_i)-\frac{1}{2\sigma^2_\nu}\sum_{j=1}^m\nu_j^2\\
    &-\frac{n+\kappa_{\sigma^2}+2}{2}\log(\sigma^2)-\frac{1}{2\sigma^2}\left(\sum_{i=1}^n\eta_i^2+\delta_{\sigma^2}\right) -\frac{n+\kappa_\psi+3}{2}\sum_{j=1}^m\log(\psi_j)-\frac{\delta_{\psi}}{2}\sum_{j=1}^m\frac{1}{\psi_j}\\
    &-\frac{1}{2\sigma^2_\lambda}\sum_{j=1}^m\frac{\lambda_j^2}{\psi_j}+\frac{\mu_\lambda}{\sigma^2_\lambda}\sum_{j=1}^m\frac{\lambda_j}{\psi_j}-\frac{\mu_\lambda^2}{2\sigma_\lambda^2}\sum_{j=1}^m\frac{1}{\psi_j}+\mbox{const},
\end{align*}
where `const' denotes terms not depending on the variables of interest. From the log-likelihood function we can derive the full conditional density functions. 
For $j=1,\ldots,m$,
\begin{align*}
    p(\nu_j\,\vert\,\mbox{rest})&\propto\exp \left[-\frac{1}{2}\left\{\nu_j^2\left(\frac{n}{\psi_j}+\frac{1}{\sigma^2_\nu}\right)-2\frac{\nu_j}{\psi_j}\sum_{i=1}^n(y_{ij}-\lambda_j\eta_i)\right\}\right]\quad\mbox{and}\\
    p(\psi_j\,\vert\,\mbox{rest})&\propto\psi_j^{-(n+\kappa_\psi+3)/2}\exp\left[-\frac{1}{2\psi_j}\left\{\sum_{i=1}^n(y_{ij}-\nu_j-\lambda_j\eta_i)^2+\frac{(\lambda_j-\mu_\lambda)^2}{\sigma^2_\lambda}+\delta_{\psi}\right\}\right].
\end{align*}
Without loss of generality, we fix $\lambda_1$ to 1 for ensuring identifiability. For $j=2,\ldots,m$,
\begin{equation*}
 p(\lambda_j\,\vert\,\mbox{rest})\propto\exp\left(-\frac{1}{2}\left[\frac{\lambda_j^2}{\psi_j}\left(\sum_{i=1}^n\eta_i^2+\frac{1}{\sigma^2_\lambda}\right)-2\frac{\lambda_j}{\psi_j}\left\{\sum_{i=1}^n (y_{ij}- \nu_j)\eta_i+\frac{\mu_\lambda}{\sigma^2_\lambda}\right\}\right]\right).
\end{equation*}
For $i=1,\ldots,n$,
\begin{equation*}
    p(\eta_i\,\vert\,\mbox{rest})\propto\exp\left[-\frac{1}{2}\left\{\eta_i^2\left(\blambda^T\diag(1/\bpsi)\blambda+\frac{1}{\sigma^2}\right)-2\eta_i\blambda^T\diag(1/\bpsi)(\by_i-\bnu)\right\}\right].
\end{equation*}
The remaining full conditional is
\begin{align*}
    p(\sigma^2\,\vert\,\mbox{rest})&\propto(\sigma^2)^{-(n+\kappa_{\sigma^2}+2)/2}\exp\left\{-\frac{1}{2\sigma^2}\left(\sum_{i=1}^n\eta_i^2+\delta_{\sigma^2}\right)\right\}.
\end{align*}

\section{Model Fitting via Mean Field Variational Bayes}\label{sec:mfvbSEM}

As for mainstream variational approximations, mean field variational Bayes boils down to finding an approximating density $q$ for which the Kullback–Leibler divergence between the approximating density itself and the posterior density function is minimized, subject to convenient restrictions on $q$.

Let $\by$ be a vector of data and $\btheta\in\Theta$ represent all model parameters. Consider an arbitrary density function $q$ defined over $\Theta$. Then the logarithm of the marginal likelihood satisfies
\begin{equation}
\log p(\by)\geq\log\underline{p}(\by;q)\equiv\int q(\btheta)\log\left\{\frac{p(\by,\btheta)}{q(\btheta)}\right\}d\btheta,
\label{eq:LowerBound}
\end{equation}
where $\underline{p}(\by;q)$ is a marginal likelihood lower-bond depending on $q$. It can be shown that maximizing this lower-bound is equivalent to minimizing the Kullback-Leibler divergence
\begin{equation*}
\mbox{KL}(q(\btheta)\,\Vert\, p(\btheta\vert\by))=\int q(\btheta)\log\left\{\frac{q(\btheta)}{p(\btheta\vert\by)}\right\}d\btheta.
\end{equation*}
If the approximating density is factorized according to a partition $(\btheta_1,\ldots,\btheta_K)$ of $\btheta$ such that $q(\btheta)=\prod_{k=1}^Kq(\btheta_k)$, then the optimal approximating densities satisfy
\begin{equation}
	q^*(\btheta_k)\propto\exp\big[E_{q(\btheta\backslash \btheta_k)}\{\log p(\btheta_k\vert\by,\btheta\backslash \btheta_k )\}\big],\quad k=1,\ldots,K,
\label{eq:qkupdate}
\end{equation}
where $E_{q(\btheta\backslash \btheta_k)}$ denotes the expectation with respect to all the approximating densities except $q(\btheta_k)$ and $\btheta\backslash \btheta_k$  represents the entries of $\btheta$ with $\btheta_k$ omitted.
It can also be shown that optimization can be performed via a coordinate ascent scheme converging to a local maximizer of the lower bound under mild regularity conditions. A more detailed introduction to MFVB is provided, for example, in Section 2.2 of \cite{ormerod2010explaining}.

Before deriving a MFVB algorithm for fitting model \eqref{eq:simpleSEM} we define some relevant notation.
For a generic parameter vector $\btheta$ of length $d$ and $f(\btheta)$ being an  elementwise function of $\btheta$, let $\bmu_{q(f(\btheta))}$ be the vector containing the expectation of the elements $f(\btheta)$ with respect to the approximating density function $q(\btheta)$. For instance, if $\btheta$ is a scalar $\theta=\sigma^2$ and $f(\sigma^2)=1/\sigma^2$, the quantity $\mu_{q(1/\sigma^2)}$ is given by:
\begin{equation*}
\mu_{q(1/\sigma^2)}\equiv\int_0^{\infty}\frac{1}{\sigma^2}q(\sigma^2)d\sigma^2.
\end{equation*}
These expressions can be obtained from a direct application of standard expectation results for exponential family distributions, such as those provided in the supplementary material of \cite{wand2017fast}.
We also define
\begin{equation*}
    \mu_{q(\theta^2)}\equiv\sigma^2_{q(\theta)} + \mu_{q(\theta)}^2,
\end{equation*}
when $q(\theta)$ is $N\big(\mu_{q(\theta)},\sigma^2_{q(\theta)}\big)$. With $\bM_{q(\bTheta^{-1})}$ we denote the expectation of the inverse of a matrix $\bTheta$ with respect to the approximating density $q(\bTheta)$. If, for example, $\bTheta$ is a $d\times d$ covariance matrix such that $\bTheta\sim\mbox{Inverse-G-Wishart}(\xi,\bLambda)$, then $\bM_{q(\bTheta^{-1})}\equiv E(\bTheta^{-1})=(\xi-d+1)\bLambda^{-1}$ (see e.g. Result 4 of Maestrini and Wand, 2021).

In order to achieve a tractable MFVB approximation for model \eqref{eq:simpleSEM}, we factorize the approximating density as follows:
\begin{align}
\begin{split}
    q(\bnu,\blambda,\bdeta,\bpsi,\sigma^2)&=q(\bnu)q(\blambda)q(\bpsi)q(\sigma^2)\prod_{i=1}^nq(\eta_i)\\
    &=q(\sigma^2)\prod_{j=1}^m\{q(\nu_j)q(\psi_j)\}\prod_{j=2}^m q(\lambda_j)\prod_{i=1}^nq(\eta_i).
\end{split}
    \label{eq:simpSEMrestr}
\end{align}
From application of (5) of \cite{ormerod2010explaining}, we have that the approximating densities expressed in \eqref{eq:simpSEMrestr} have the following optimal forms:
\begin{equation*}
    \begin{array}{c}
    q^*(\nu_j)\,\,\,\mbox{is}\,\,\,N\big(\mu_{q(\nu_j)},\sigma^2_{q(\nu_j)}\big),\,\,\,j=1,\ldots,m,\\[1ex]
    \mbox{with}\,\,\,\mu_{q(\nu_j)}\equiv \sigma^2_{q(\nu_j)}\mu_{q(1/\psi_j)}\sum_{i=1}^n(y_{ij}-\mu_{q(\lambda_j)}\mu_{q(\eta_i)})\,\,\,\mbox{and}\,\,\,\sigma^2_{q(\nu_j)}\equiv1\big/(n\mu_{q(1/\psi_j)}+1/\sigma^2_\nu);
    \end{array}
\end{equation*}
\begin{equation*}
    \begin{array}{c}
    q^*(\lambda_j)\,\,\,\mbox{is}\,\,\,N\big(\mu_{q(\lambda_j)},\sigma^2_{q(\lambda_j)}\big),\,\,\,j=2,\ldots,m,\\[1ex]
    \mbox{with}\,\,\,\mu_{q(\lambda_j)}\equiv\sigma^2_{q(\lambda_j)}\left[\sum_{i=1}^n\big\{\mu_{q(\eta_i)}(y_{ij}-\mu_{q(\nu_j)})\big\}+\mu_\lambda/\sigma^2_\lambda\right]\mu_{q(1/\psi_j)}\,\,\,
    \mbox{and}\\[1ex]
    \sigma^2_{q(\lambda_j)}\equiv1\big/\big\{\mu_{q(1/\psi_j)}\big(\sum_{i=1}^n\mu_{q(\eta_i^2)}+1/\sigma^2_{\lambda}\big)\big\};
    \end{array}
\end{equation*}
\begin{equation*}
    \begin{array}{c}
    q^*(\eta_i)\,\,\,\mbox{is}\,\,\,N\big(\mu_{q(\eta_i)},\sigma^2_{q(\eta_i)}\big),\,\,\,i=1,\ldots,n,\\[1ex]
    \mbox{with}\,\,\,\mu_{q(\eta_i)}\equiv\sigma^2_{q(\eta_i)}\sum_{j=1}^m\big\{\mu_{q(1/\psi_j)}\mu_{q(\lambda_j)}(y_{ij}-\mu_{q(\nu_j)})\big\}\,\,\,
    \mbox{and}\\[1ex]
    \sigma^2_{q(\eta_i)}\equiv1\big/\big(\sum_{j=1}^m\mu_{q(1/\psi_j)}\mu_{q(\lambda_j^2)}+\mu_{q(1/\sigma^2)}\big);
    \end{array}
\end{equation*}
\begin{equation*}
    \begin{array}{c}
    q^*(\psi_j)\,\,\,\mbox{is}\,\,\,\mbox{Inverse-}\chi^2\big(\kappa_{q(\psi_j)},\delta_{q(\psi_j)}),\,\,\,j=1,\ldots,m,\\[1ex]
    \mbox{with}\,\,\,\kappa_{q(\psi_j)}\equiv n+\kappa_\psi+1\,\,\,
    \mbox{and}\\[1ex]\delta_{q(\psi_j)}\equiv\sum_{i=1}^n\big(y_{ij}^2+\mu_{q(\nu_j^2)}+\mu_{q(\lambda_j^2)}\mu_{q(\eta_i^2)}-2y_{ij}\mu_{q(\nu_j)}-2y_{ij}\mu_{q(\lambda_j)}\mu_{q(\eta_i)}+2\mu_{q(\nu_j)}\mu_{q(\lambda_j)}\mu_{q(\eta_i)}\big)\\[1ex]
    +\frac{1}{\sigma^2_\lambda}\big(\mu_{q(\lambda_j^2)}-2\mu_{\lambda}\mu_{q(\lambda_j)}+\mu^2_\lambda\big)+\delta_{\psi};
    \end{array}
\end{equation*}
\begin{equation*}
    \begin{array}{c}
    q^*(\sigma^2)\,\,\,\mbox{is}\,\,\,\mbox{Inverse-}\chi^2\big(\kappa_{q(\sigma^2)},\delta_{q(\sigma^2)}\big),\\[1ex]
    \mbox{with}\,\,\,\kappa_{q(\sigma^2)}\equiv n+\kappa_{\sigma^2}\,\,\,\mbox{and}\,\,\,\delta_{q(\sigma^2)}\equiv\sum_{i=1}^n\mu_{q(\eta_i^2)}+\delta_{\sigma^2}.
    \end{array}
\end{equation*}

A MFVB scheme for fitting model \eqref{eq:simpleSEM} under restriction \eqref{eq:simpSEMrestr} is listed as Algorithm \ref{alg:MFVBsimpSEM}, which also provides expressions for $\mu_{q(\nu_j^2)}$, $\mu_{q(\lambda_j^2)}$, $\mu_{q(1/\psi_j)}$, $j=1,\ldots,m$, $\mu_{q(\eta_i^2)}$, $i=1,\ldots,n$, and $\mu_{q(1/\sigma^2)}$. The algorithm outputs are the parameters of the optimal approximating densities. We refer to this scheme as the basic MFVB algorithm.
\begin{algorithm}[!th]
	\begin{center}
		\begin{minipage}[t]{154mm}
			\begin{small}
				\begin{itemize}
					\setlength\itemsep{4pt}
					\item[] \textbf{Data Input:} $\by_i$, $i=1,\ldots,n$, vectors of length $m$.
					\item[] \textbf{Hyperparameter Input:} $\mu_{\lambda}\in\mathbb{R}$ and $\sigma_\nu,\sigma_\lambda,\kappa_\psi,\delta_\psi,\kappa_{\sigma^2},\delta_{\sigma^2}\in\mathbb{R}^+$.
					\item[] \textbf{Initialize:} $\mu_{q(\nu_j)}$, $\mu_{q(\nu_j^2)},\mu_{q(1/\psi_j)}\in\mathbb{R}^+$, $j=1,\ldots,m$; $\mu_{q(\lambda_j)}$, $\mu_{q(\lambda_j^2)}\in\mathbb{R}^+$, $j=2,\ldots,m$; $\mu_{q(\eta_i)}$, $\mu_{q(\eta_i^2)}\in\mathbb{R}^+$, $i=1,\ldots,n$; $\mu_{q(1/\sigma^2)}\in\mathbb{R}^+$.
					\item[] \textbf{Set:}  $\mu_{q(\lambda_1)}=\mu_{q(\lambda_1^2)}=1$; $\kappa_{q(\sigma^2)}=n+\kappa_{\sigma^2}$; $\kappa_{q(\psi_j)} = n+\kappa_\psi+1$, $j = 1,\dots,m$.
					\item[] \textbf{Cycle until convergence:}
					\begin{itemize}
						\setlength\itemsep{4pt}
						\item[] For $j=1,\ldots,m$:
						\begin{itemize}
						    \setlength\itemsep{4pt}
    					    \item[] $\sigma^2_{q(\nu_j)}\longleftarrow1\big/(n\mu_{q(1/\psi_j)}+1/\sigma^2_\nu)$
    					    \item[] $\mu_{q(\nu_j)}\longleftarrow \sigma^2_{q(\nu_j)}\mu_{q(1/\psi_j)}\sum_{i=1}^n(y_{ij}-\mu_{q(\lambda_j)}\mu_{q(\eta_i)})$
    					    \item[] $\mu_{q(\nu_j^2)}\longleftarrow\sigma^2_{q(\nu_j)}+\mu_{q(\nu_j)}^2$
    					    \item[] $\delta_{q(\psi_j)}\longleftarrow\sum_{i=1}^n\big(y_{ij}^2+\mu_{q(\nu_j^2)}+\mu_{q(\lambda_j^2)}\mu_{q(\eta_i^2)}-2y_{ij}\mu_{q(\nu_j)}-2y_{ij}\mu_{q(\lambda_j)}\mu_{q(\eta_i)}$
    					    \item[] $\qquad\qquad\qquad+2\mu_{q(\nu_j)}\mu_{q(\lambda_j)}\mu_{q(\eta_i)}\big)+\frac{1}{\sigma^2_\lambda}\big(\mu_{q(\lambda_j^2)}-2\mu_{\lambda}\mu_{q(\lambda_j)}+\mu^2_\lambda\big)+\delta_{\psi}$
    					    \item[] $\mu_{q(1/\psi_j)}\longleftarrow \kappa_{q(\psi_j)}/\delta_{q(\psi_j)}$
    					   	\item[] If $j>1$:
						       \begin{itemize}
						        \setlength\itemsep{4pt}
    					            \item[]  $\sigma^2_{q(\lambda_j)}\longleftarrow1\big/\big\{\mu_{q(1/\psi_j)}\big(\sum_{i=1}^n\mu_{q(\eta_i^2)}+1/\sigma^2_{\lambda}\big)\big\}$
    					            \item[] $\mu_{q(\lambda_j)}\longleftarrow\sigma^2_{q(\lambda_j)}\left[\sum_{i=1}^n\big\{\mu_{q(\eta_i)}(y_{ij}-\mu_{q(\nu_j)})\big\}+\mu_\lambda/\sigma^2_\lambda\right]\mu_{q(1/\psi_j)}$
    					            \item[] $\mu_{q(\lambda_j^2)}\longleftarrow\sigma^2_{q(\lambda_j)}+\mu_{q(\lambda_j)}^2$
    					       \end{itemize}
    			        \end{itemize}
						\item[] For $i=1,\ldots,n$:
						\begin{itemize}
						    \setlength\itemsep{4pt}
    					    \item[] $\sigma^2_{q(\eta_i)}\longleftarrow1\big/\big(\sum_{j=1}^m\mu_{q(1/\psi_j)}\mu_{q(\lambda_j^2)}+\mu_{q(1/\sigma^2)}\big)$
                            \item[] $\mu_{q(\eta_i)}\longleftarrow\sigma^2_{q(\eta_i)}\sum_{j=1}^m\big\{\mu_{q(1/\psi_j)}\mu_{q(\lambda_j)}(y_{ij}-\mu_{q(\nu_j)})\big\}$
                            \item[] $\mu_{q(\eta_i^2)}\longleftarrow\sigma^2_{q(\eta_i)}+\mu_{q(\eta_i)}^2$
    			        \end{itemize}
    			        \item[] $\delta_{q(\sigma^2)}\longleftarrow\sum_{i=1}^n\mu_{q(\eta_i^2)}+ \delta_{\sigma^2}$
    			        \item[] $\mu_{q(1/\sigma^2)}\longleftarrow\kappa_{q(\sigma^2)}/\delta_{q(\sigma^2)}$
					\end{itemize}
					\item[] \textbf{Output:} $\mu_{q(\nu_j)}$, $\sigma^2_{q(\nu_j)}$, $\mu_{q(\lambda_j)}$, $\sigma^2_{q(\lambda_j)}$, $\kappa_{q(\psi_j)}$, $\delta_{q(\psi_j)}$,   $j=1,\ldots,m$; $\mu_{q(\eta_i)}$, $\sigma^2_{q(\eta_i)}$, $i=1,\ldots,n$; $\kappa_{q(\sigma^2)}$, $\delta_{q(\sigma^2)}$.
				\end{itemize}
			\end{small}
		\end{minipage}
	\end{center}
	\caption{\textit{Algorithm for fitting model \eqref{eq:simpleSEM} via MFVB, under restriction \eqref{eq:simpSEMrestr}.}}
	\label{alg:MFVBsimpSEM}
\end{algorithm}

\section{Illustration on Holzinger \& Swineford (1939) Data}\label{sec:realDataNoBoot}

\begin{figure}[!ht]
	\centering
	{\includegraphics[width=0.5\textwidth]{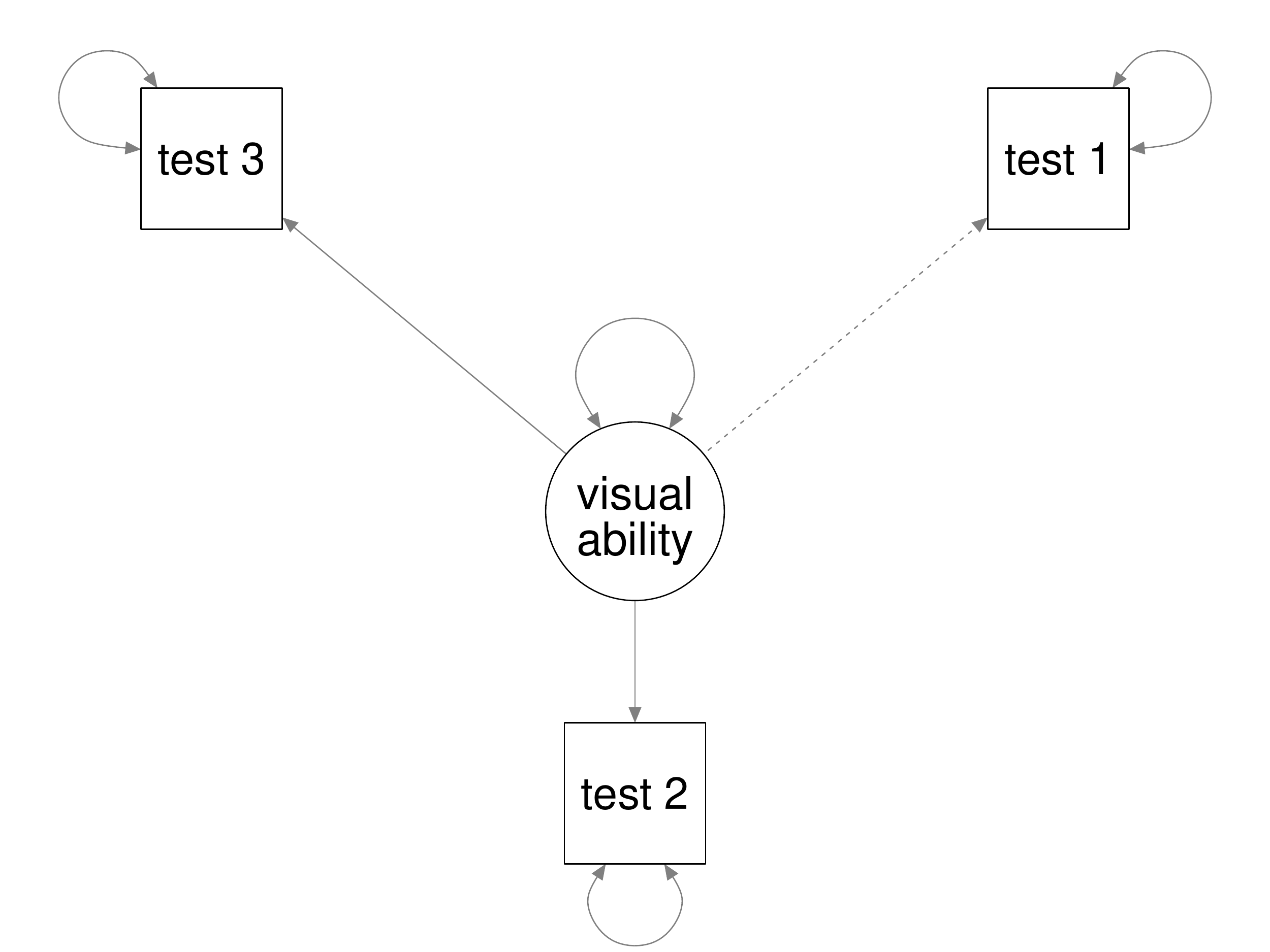}}
	\caption{\it Path diagram of model \eqref{eq:simpleSEM} adapted to the Holzinger \& Swineford (1939) data, from which we consider $m=3$ outcomes. Squares depict observable outcomes (tests 1, 2 and 3). The unobservable component, visual ability, is represented by a circle. Circular arrows represent correlations, whereas the dashed line indicates that the loading corresponding to test 1 was fixed to 1 for identifiability.}
	\label{fig:HS1939_PathDiagram} 
\end{figure}

In this section, we consider a well-known dataset from \cite{holzinger1939study} and demonstrate benefits and limitations of the basic MFVB algorithm for fitting a simple SEM. The full original dataset, available in the \textsf{R} \citep{R2021} package \textit{MBESS} \citep{MBESSpackage}, consists of mental ability test scores of 301 seventh- and eighth-grade students from two schools (Pasteur and Grant-White). This dataset includes scores from 26 tests for measuring the participants' spatial, verbal, mental speed, memory and mathematical abilities. We utilized a cleaned version of this dataset from the \textsf{R} package \textit{lavaan}. The data in \textit{lavaan} has been rescored through ratio transformations so that the transformed scores are within a desired range. For illustration purposes, we only used outcomes from the first three tests: the \textit{visual perception}, \textit{cubes} and \textit{lozenges} tests. These outcomes are hypothesized to be associated with the participants' visual ability.
The model for the examined dataset is represented by the path diagram labeled as Figure \ref{fig:HS1939_PathDiagram}.

We fitted the confirmatory factor analysis model \eqref{eq:simpleSEM} to this data using MFVB and compared the results to MCMC, which we consider our benchmark. Both methods were implemented in \textsf{R} and MFVB was run through Algorithm \ref{alg:MFVBsimpSEM} initialized at suitable points of the parameter space. The algorithm was stopped after the relative error between optimal density parameter estimates from two consecutive iterations went below 0.01. MCMC was performed via the \textsf{R} package \textsf{rstan} \citep{RStan2020}. We ran 15,000 iterations of MCMC and discarded the first 7,500 as burn-in. MFVB converged in 0.1 seconds 
after 98 iterations, while MCMC took 307
seconds. In addition, we implemented variational inference using automated differentiation variational inference (ADVI) in \textsf{Stan} \citep[e.g.][]{kucukelbir2015}. However, we found the approximation to be less accurate than the one offered by our algorithm, therefore we do not provide details on this approach in our examples. 

\begin{figure}[!ht]
	\centering
	{\includegraphics[width=1\textwidth]{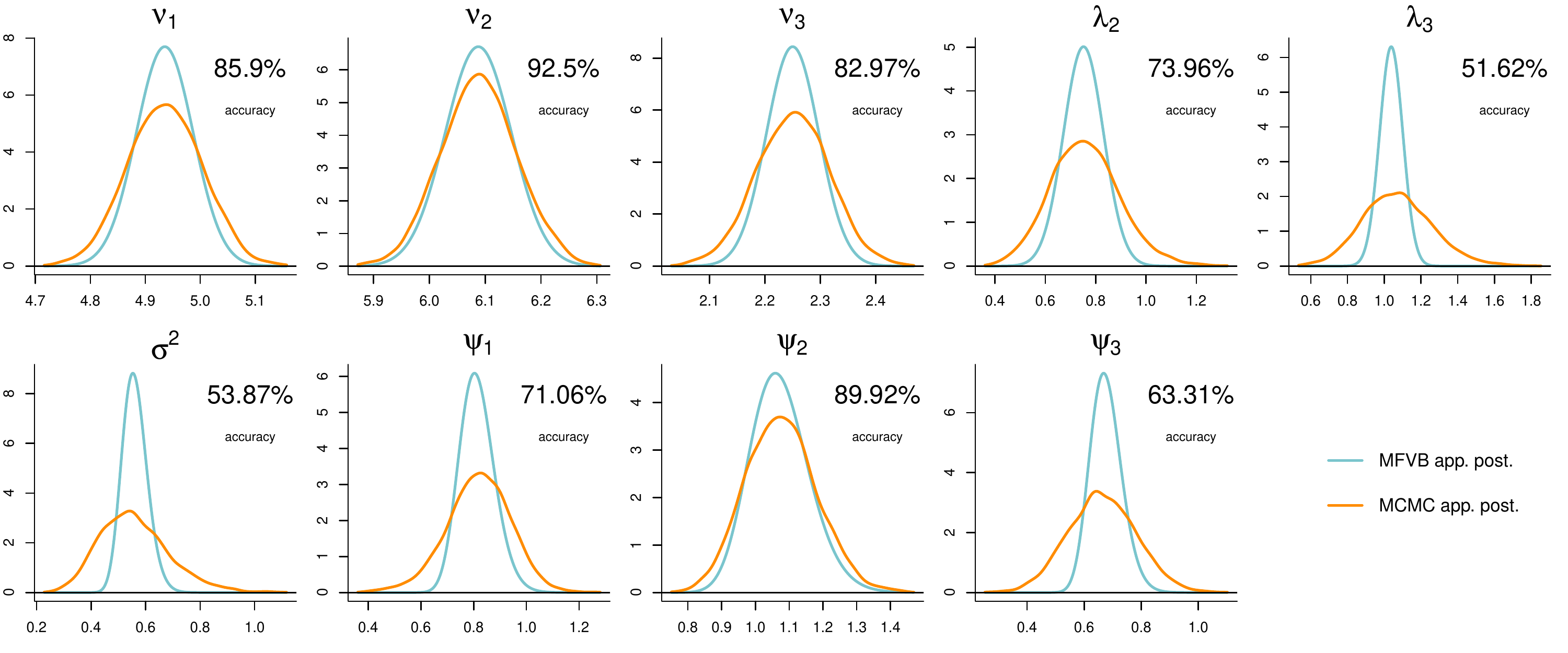}}
	\caption{\it Approximate marginal posterior densities of the parameters of interests from model \eqref{eq:simpleSEM} fitted to the examined subset of the Holzinger \& Swineford (1939) data. The curves are obtained via MFVB (light blue) and MCMC (orange). The accuracy displayed in each plot is calculated using \eqref{eq:accExpr}.}
	\label{fig:HS1939_noboot} 
\end{figure}

Figure \ref{fig:HS1939_noboot} shows the approximate marginal posterior densities of the parameters of interest from MFVB and MCMC. 
The accuracy of the MFVB approximation is also displayed as a percentage value, with $100\%$ indicating perfect matching between the MFVB and MCMC approximate posterior density functions. 
For a single parameter $\theta$, the accuracy of the approximation of a density $q(\theta)$ to the posterior density $p(\theta\vert\by)$ is computed as
\begin{equation}
\mbox{accuracy}\equiv 100\left(1-\frac{1}{2}\int_{-\infty}^{\infty}\big\vert q(\theta)-p(\theta\vert\by)\big\vert d\theta\right)\%.
\label{eq:accExpr}
\end{equation}
In practice, we obtain this quantity by replacing $p(\theta\vert\by)$ in \eqref{eq:accExpr} with the MCMC density estimates of the posterior density functions. 

Taking MCMC as a benchmark, it is clear that MFVB can estimate the posterior means fairly accurately. However, MFVB underestimates the posterior variance of all the parameters, in particular of $\lambda_3$, $\sigma^2$ and $\psi_3$, which are also associated with lower accuracy values. This is a common issue of MFVB related to the mean field restriction imposed to the approximating density  \citep{titterington2004bayesian,wang2005inadequacy}. In the next section, we propose a solution to improve posterior variance estimates based on bootstrap.

%

\section{Improved Variational Inference}\label{sec:bootrapMethod}

The output of Algorithm \ref{alg:MFVBsimpSEM} can be directly used to produce credible intervals for the model parameters of interest. However, Figure \ref{fig:HS1939_noboot} foretells these confidence intervals may present poor coverage performances due to the bias and variance underestimation issues of MFVB. Bootstrap can be the solution to these issues \citep{chen2018use}.
Usage of bootstrap in conjunction with variational approximations has not been widely explored in the literature, despite this technique may substantially improve variational inference performances, as shown in our real and simulated analyses.

In this section we study the use of nonparametric bootstrap for improving the accuracy of MFVB approximations for SEM parameters. The idea behind nonparametric bootstrap approaches is to sample with replacement from the original dataset, recompute the variational estimates of parameters for each bootstrap sample and use the distribution of these estimates or other quantities related to them to derive uncertainty measures. \cite{chen2018use} discuss asymptotic normality and validity of nonparametric bootstrap for variational approximate maximum likelihood estimators and also affirm these results apply to Bayesian variational estimators when the prior is sufficiently smooth. Therefore we follow the prescriptions from this reference for constructing more accurate credible intervals and replace the variational approximate maximum likelihood estimates with point estimates obtained via MFVB.  

The first issue arising when designing a bootstrap strategy is the choice of an appropriate sampling procedure. Model \eqref{eq:simpleSEM} and the SEMs examined in this manuscript can be interpreted as models from grouped data, where groups correspond to individuals.
In \citet[Section 3.8]{davison1997bootstrap} two nonparametric sampling strategies for hierarchical or multilevel data are examined through a basic problem involving observations that are part of mutually exclusive groups. Both strategies involve randomly sampling groups with replacement at a first stage and then randomly sampling within the groups selected at the first stage. The first strategy consists of sampling all the group units without replacement at the second stage, which implies simply keeping the selected groups intact after the first step. The second strategy corresponds to sampling with replacement at the second stage. \cite{davison1997bootstrap} show that the first strategy has to be preferred, since it more closely mimics the variation properties of the data.
Thus, we follow this advice and simply sample with replacement the $\by_i$ vectors of observed data, maintaining the sampled vector referring to individual $i$ unchanged.
We then fit the model to the bootstrapped dataset using MFVB. This procedure is repeated $B$ times and  we use the final set of bootstrap MFVB parameter estimates to produce credible intervals. 

There exist several ways of constructing bootstrap confidence intervals (see for example \citeauthor{hall2013bootstrap}, \citeyear{hall2013bootstrap}). 
\cite{chen2018use} focus on two common nonparametric approaches for variational inference, the \textit{percentile} and \textit{(studentized) pivotal} methods, and also caution against use of parametric bootstrap, since variational estimators usually do not coincide with maximum likelihood estimators. Percentile bootstrap confidence intervals can be obtained simply through the percentiles of the bootstrap distribution of the variational estimator. The construction of credible intervals for variational inference using the percentile approach was first explored in \cite{wang2017variational}.
The (studentized) pivotal approach \citep{wasserman2006all} may produce confidence intervals with a higher-order correctness if a consistent estimator of the variance of the estimator is used \citep{hall2013bootstrap}.

The main steps to compute percentile and pivotal bootstrap credible intervals in conjuction with variational approximations for a generic SEM parameter $\theta$ are the following:
\begin{itemize}
    \item Find the variational inference estimate $\hat{\theta}_{\tiny\mbox{VI}}$ of $\theta$ using the original dataset; for the pivotal approach, also find an estimate of the variance of the variational inference estimator, $\hat{\sigma}_{\tiny\mbox{VI}}$.
    \item For $b=1,\ldots,B$:
    \begin{itemize}
        \item Sample with replacement $n$ vectors from $\by_1,\ldots, \by_n$ to form the $b$th bootstrap dataset.
        \item Find the variational inference estimate $\hat{\theta}_{\tiny\mbox{B-VI}}^{(b)}$ of $\theta$ using the $b$th bootstrap dataset.
        \item If using the pivotal approach, also find an estimate of the variance of the variational inference estimator, $\hat{\sigma}_{\tiny\mbox{B-VI}}^{(b)}$.
        \item Calculate $\delta_{\tiny\mbox{B-VI}}^{(b)}=\hat{\theta}_{\tiny\mbox{B-VI}}^{(b)}-\hat{\theta}_{\tiny\mbox{VI}}$.
        \item If using the pivotal approach, also calculate $\tau^{(b)}_{\tiny\mbox{B-VI}}=\delta_{\tiny\mbox{B-VI}}^{(b)}/\hat{\sigma}_{\tiny\mbox{B-VI}}^{(b)}$.
    \end{itemize}
    \item Given a credible level $\alpha$:
    \begin{itemize}
        \item If using the percentile approach, compute the $\alpha/2$ and  $1-\alpha/2$ quantiles of the empirical distribution of $\delta_{\tiny\mbox{B-VI}}^{(1)},\ldots,\delta_{\tiny\mbox{B-VI}}^{(B)}$, $q_{\alpha/2}^{per}$ and $q_{1-\alpha/2}^{per}$.
        \item If using the pivotal approach, compute the $1-\alpha/2$ quantile of the empirical distribution of $|\tau_{\tiny\mbox{B-VI}}^{(1)}|,\ldots,|\tau_{\tiny\mbox{B-VI}}^{(B)}|$, $q_{1-\alpha/2}^{piv}$.
    \end{itemize} 
    \item Compute the credible interval:
    \begin{itemize}
        \item If using the percentile approach, the credible interval is given by:
        \begin{equation*}
            \big[\hat{\theta}_{\tiny\mbox{VI}}+q_{\alpha/2}^{per},\,\,\,\hat{\theta}_{\tiny\mbox{VI}}+q_{1-\alpha/2}^{per}\big].
        \end{equation*}
        \item If using the pivotal approach, the credible interval is given by:
        \begin{equation*}
            \big[\hat{\theta}_{\tiny\mbox{VI}}-\hat{\sigma}_{\tiny\mbox{VI}}q_{1-\alpha/2}^{piv},\,\,\,\hat{\theta}_{\tiny\mbox{VI}}+\hat{\sigma}_{\tiny\mbox{VI}}q_{1-\alpha/2}^{piv}\big].
        \end{equation*}
    \end{itemize}
\end{itemize}
In our illustrations, $\hat{\theta}_{\tiny\mbox{VI}}$ and $\hat{\theta}_{\tiny\mbox{B-VI}}^{(b)}$ correspond to the mean of the MFVB approximating density. From Section \ref{sec:mfvbSEM}, for instance, the MFVB approximating density for the parameter $\nu_j$ of model \eqref{eq:simpleSEM} is Normal with mean  $\mu_{q(\nu_j)}$, therefore the estimates $\hat{\theta}_{\tiny\mbox{VI}}$ and $\hat{\theta}_{\tiny\mbox{B-VI}}^{(b)}$ of $\theta=\nu_j$ are simply the optimal $\mu_{q(\nu_j)}$ value obtained at convergence of Algorithm \ref{alg:MFVBsimpSEM}. For $\sigma^2$, and similarly for the parameters of model \eqref{eq:simpleSEM} whose approximating density is Inverse-$\chi^2$, these estimates are given by $\delta_{q(\sigma^2)}/(\kappa_{q(\sigma^2)}-2)$.

\begin{figure}[!ht]
	\centering
	{\includegraphics[width=1\textwidth]{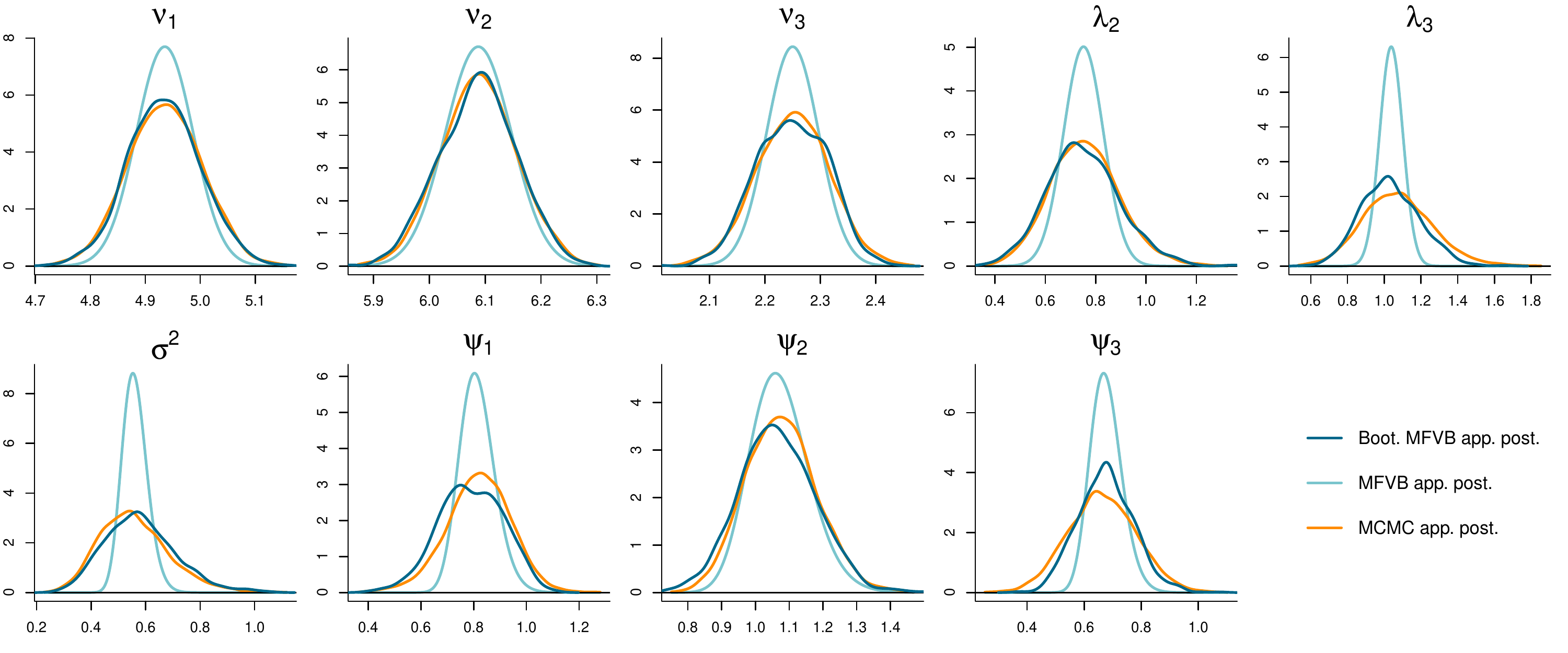}}
	\caption{\it The approximating densities of Figure \ref{fig:HS1939_noboot} together with density estimates obtained using MFVB point estimates from 1,000 bootstrap samples (dark blue curves).}
	\label{fig:HS1939_Boot} 
\end{figure}

We repeated the analysis of the Holzinger \& Swineford (1939) data using 1,000 bootstrap samples to demonstrate the effect of bootstrap on the variance of the variational inference approximate posterior densities. Figure \ref{fig:HS1939_Boot} shows the plots of Figure \ref{fig:HS1939_noboot} with the addition of dark blue curves which represent density estimates obtained from the set of bootstrap point estimates $\hat{\theta}_{\tiny\mbox{B-VI}}^{(1)},\ldots,\hat{\theta}_{\tiny\mbox{B-VI}}^{(B)}$, for each parameter of interest. These curves show that bootstrap significantly improves the accuracy of the MFVB approximate marginal posterior densities when compared to the original MFVB approximations. 

%

\section{Simulated Data Study}\label{sec:simStudy}

In this section we provide a simulation exercise to examine the effect of data resampling on the coverage performances of MFVB credible intervals for the simple SEM parameters of interest. 

We generated 1,000 datasets using the MCMC point estimates of the parameters of model \eqref{eq:simpleSEM} fitted using the Holzinger \& Swineford data described in Section \ref{sec:realDataNoBoot}. The parameter point estimates were obtained by averaging over the retained MCMC chains of each parameter. For each of the 1,000 simulated datasets we ran MCMC and simple MFVB, and implemented the bootstrap procedure outlined in Section \ref{sec:bootrapMethod} with $B$ equal to 100, 500 or 1,000. We then constructed $95\%$ percentile and pivotal credible intervals for each parameter of interest. The pivotal credible intervals were based on estimates $\hat{\sigma}_{\tiny\mbox{B-VI}}^{(b)}$ of the variance of the variational inference estimator given by the variance of the MFVB approximating densities. These correspond, for instance, to $\sigma^2_{q(\nu_j)}$ for $\nu_j$ or $2\delta_{q(\sigma^2)}^2/\{(\kappa_{q(\sigma^2)}-2)^2(\kappa_{q(\sigma^2)}-4)\}$ for $\sigma^2$, since $q^*(\nu_j)$ is N$\big(\mu_{q(\nu_j)},\sigma^2_{q(\nu_j)}\big)$ and $q^*(\sigma^2)$ is Inverse-$\chi^2\big(\kappa_{q(\sigma^2)},\delta_{q(\sigma^2)}\big)$.  Credible intervals from simple MFVB without bootstrap were also obtained. These were computed from the $2.5\%$ and $97.5\%$ quantiles of the MFVB approximating densities. In addition, we produced jackknife credible intervals \citep{efron1982jackknife}. Jackknife may be considered an approximation to bootstrap that can be implemented at a lower computational cost.
For $i=1,\ldots,n$, we computed the variational estimate $\hat{\theta}_{\tiny\mbox{J-VI}}^{(i)}$ of $\theta$ using the $i$th jackknife dataset generated by omitting $\by_i$ from the original dataset. The $95\%$ jackkife confidence intervals were obtained using the quantiles of a standard normal distribution and the jackknife mean and standard error estimates respectively given by  
$$\hat{\theta}_{\tiny\mbox{J-VI}}^{(\cdot)}\equiv\sum_{i=1}^n\hat{\theta}_{\tiny\mbox{J-VI}}^{(i)}\quad\mbox{and}\quad\hat{\sigma}_{\tiny\mbox{J-VI}}\equiv\left\{\frac{n-1}{n}\sum_{i=1}^n\big(\hat{\theta}_{\tiny\mbox{J-VI}}^{(i)}-\hat{\theta}_{\tiny\mbox{J-VI}}^{(\cdot)}\big)^2\right\}^{1/2}.$$
MCMC implemented in \textsf{rstan} was again used as a benchmark. The MCMC credible intervals were given by the $2.5\%$ and $97.5\%$ quantiles of 7,500 retained MCMC iterations following a burn-in of equal length. The simulation study was implemented on a personal computer with a 64 bit Windows 10 operating system, an Intel i7-7500U central processing unit at 2.7 gigahertz and 16 gigabytes of random access memory.

The computational times of simple MFVB and MCMC from the simulation study are reported in Table \ref{tab:SimStudCompTimes}, which shows MFVB was several orders of magnitude faster than MCMC. The median of the ratios between MCMC and MFVB running times was greater than 5,000. This indicates that MFVB would still provide accurate inference at reduced time even when running 1,000 bootstrap iterations in a non-parallelized system.

\begin{table}[!ht]
\centering
\begin{tabular}{l c c c}
& 1st quartile & median & 3rd quartile \\ \hline
MFVB & 0.038 & 0.048 & 0.060 \\
MCMC & 205.3 & 240.5 & 287.4 \\
\hline
\end{tabular}
\caption{Computational times in seconds of MFVB (no bootstrap) and MCMC from the simulation study.}
\label{tab:SimStudCompTimes}
\end{table}

\begin{table}[!ht]
\resizebox{\textwidth}{!}{
\begin{tabular}{l l l l l l l l l l}
& $\nu_1$ & $\nu_2$ & $\nu_3$ & $\lambda_2$ & $\lambda_3$  & $\sigma^2$   & $\psi_1$ & $\psi_2$ & $\psi_3$ \\ \hline
\textbf{MFVB} & \textbf{0.851} & \textbf{0.914} & \textbf{0.839} & \textbf{0.776} & \textbf{0.543} & \textbf{0.552} & \textbf{0.735} & \textbf{0.930} & \textbf{0.679}\\
MFVB with jackknife & 0.940 & 0.943 & 0.939 & 0.951 & 0.890 & 0.949 & 0.943 & 0.970 & 0.935 \\
MFVB with per. boot. (100)  & 0.917 & 0.926 & 0.922 & 0.936 & 0.894 & 0.913 & 0.895 & 0.950 & 0.922 \\
MFVB with per. boot. (500) & 0.938 & 0.945 & 0.940 & 0.955 & 0.904 & 0.936 & 0.918 & 0.959 & 0.937 \\
\textbf{MFVB with per. boot. (1,000)} & \textbf{0.941} & \textbf{0.947} & \textbf{0.940} & \textbf{0.957} & \textbf{0.905} & \textbf{0.938} & \textbf{0.925} & \textbf{0.958} & \textbf{0.940} \\
MFVB with piv. boot. (100) &  0.966 & 0.968 & 0.956 & 0.958 & 0.907 & 0.972 & 0.980 & 0.983 & 0.966 \\
MFVB with piv. boot. (500) & 0.971 & 0.972 & 0.971 & 0.967 & 0.913 & 0.981 & 0.989 & 0.990 & 0.975 \\
MFVB with piv. boot. (1,000) & 0.972 & 0.974 & 0.972 & 0.968 & 0.920 & 0.980 & 0.990 & 0.991 & 0.975 \\ 
\textbf{MCMC} & \textbf{0.941} & \textbf{0.946} & \textbf{0.939} & \textbf{0.955} & \textbf{0.944} & \textbf{0.949} & \textbf{0.944} & \textbf{0.968} & \textbf{0.952} \\
\hline
\end{tabular}}
\caption{Average empirical coverage percentages for advertized 95\% credible intervals of the parameters of interest from the simulation study described in Section \ref{sec:simStudy} based on 1,000 replications. Model \eqref{eq:simpleSEM} was fitted via MFVB using Algorithm \ref{alg:MFVBsimpSEM}, MFVB in conjunction with jackknife and bootstrap resamplings, and  Markov chain Monte Carlo (MCMC) in \textsf{rstan}. For the percentile and pivotal bootstrap results, 100, 500 and 1,000 bootstrap iterations were used.}
\label{tab:SimStudRes}
\end{table}

Table \ref{tab:SimStudRes} displays the empirical coverage percentages from the simulation study. For each parameter, the percentage of coverage is calculated as the proportion of simulations where the true parameter falls inside a 95\% credible interval produced via MCMC, MFVB or MFVB paired with a resampling strategy. Different bootstrap results are shown according to the number of bootstrap samples (100, 500 and 1,000) and the approach (percentile and pivotal) used to create the parameter credible intervals.
The empirical coverages of all credible intervals obtained via MFVB together with both jackknife and bootstrap resampling are closer to the 95\% nominal level than simple MFVB. MCMC empirical coverages seem to be around the advertised 95\% level. Jackknife provided satisfactory results, whereas the pivotal bootstrap credible intervals produced excessive coverage. This is clearly noticeable when the MFVB and MCMC credible intervals are plotted together.

\begin{figure}[!htp]
	\centering
	{\includegraphics[width=1\textwidth]{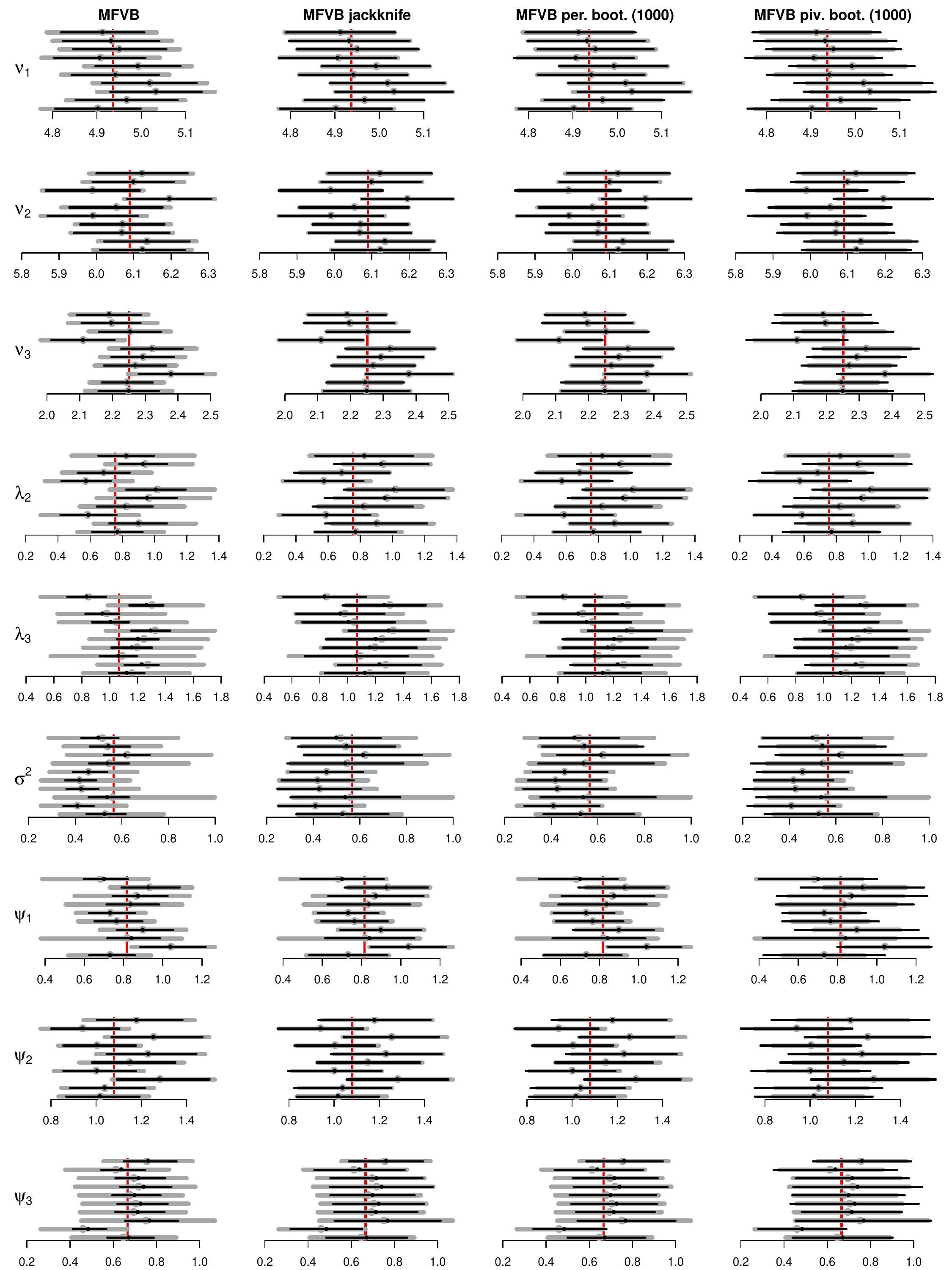}}
	\caption{\it Representation of a selection of 95\% credible intervals from the simulation study described in Section \ref{sec:simStudy}. Each plot refers to a parameter of interest and method, and displays credible intervals of 10 randomly selected simulated datasets from the 1,000 simulation study replicates. The thick gray lines are MCMC credible intervals. The black lines are MFVB (1st column), MFVB with jackknife (2nd column), MFVB percentile bootstrap (3rd column) and MFVB pivotal bootstrap (4th colum) credible intervals. The percentile and pivotal bootstrap credible intervals were obtained from 1,000 bootstrap datasets for each simulation study replicate. The red vertical lines indicate the true parameters with which the simulated datasets were generated.}
	\label{fig:Sim_CI} 
\end{figure}

Figure \ref{fig:Sim_CI} shows plots with credible intervals for each parameter of interest obtained via MCMC, simple MFVB or MFVB paired with resampling (jacknife and percentile or pivotal bootstrap based on 1,000 bootstrapped datasets) from 10 randomly selected simulated datasets. The thick gray lines are MCMC credible intervals and black lines are used for the credible intervals produced with all the other strategies based on MFVB. From these plots it is evident that resampling allows to obtain wider credible intervals that better mimic the MCMC benchmark, although the pivotal approach tends to overestimate the variance. This may be due to the fact that the pivotal credible intervals have been produced using inconsistent estimators of the variance of the variational inference parameter estimators. 

%

\section{A Model with Multiple Factors}\label{sec:extension}

In this section we extend our framework and develop MFVB for data with multiple latent factors. With a real data application, we then illustrate the performances of MFVB supported by bootstrap.

Sticking to the prescriptions of \cite{lee2007structural}, we examine the following model for multiple latent factors:
\begin{equation}
    \begin{array}{c}
        \by_{i}\,\vert\,\bnu,\bLambda,\bdeta_{i},\bpsi\simind N\big(\bnu+\bLambda\bdeta_i,\diag(\bpsi)\big),\,\,\,\bdeta_i\,\vert\,\bSigma\simind N(\bzero,\bSigma),\,\,\, i=1,\ldots,n,\\[1ex]
        \bLambda_{j}^T\,\vert\,\psi_{j}\simind N(\bmu_{\bLambda},\psi_{j}\bSigma_{\bLambda}),\,\,\, \nu_{j}\simind N(0,\sigma^2_\nu),\,\,\,\psi_{j}\simind\mbox{Inverse-}\chi^2(\kappa_{\psi},\delta_{\psi}),\,\,\, j=1,\ldots,m,\\[1.2ex]     \bSigma\sim\mbox{Inverse G-Wishart}(\Gfull,\xi_{\bSigma},\bLambda_{\bSigma}).
    \end{array}
\label{eq:complexSEM}
\end{equation}
The vector $\by_{i}=[(\by_i)_{11},\ldots,(\by_i)_{1m_1},\ldots,(\by_i)_{p1},\ldots,(\by_i)_{pm_p}]$ contains $m=\sum_{k=1}^p m_k$ observed outcomes and follows a Normal distribution with diagonal covariance matrix. Here $(\by_i)_{kj'}$ denotes the $j'$th observed outcome, $j'=1,\ldots,m_k$, of the latent factor $k$, $k=1,\ldots,p$, for the $i$th individual, $i=1,\ldots,n$. The vector of intercepts $\bnu$ has length $m$, $\bLambda$ is a matrix of factor loadings with size $m\times p$ and $\bdeta_i$ is a $p\times 1$ vector of unobserved factors which is normally distributed with $p\times p$ covariance matrix $\bSigma$. Also, $\bLambda_j^T$ denotes the $j$th row of $\bLambda$ and follows a Normal distribution with covariance matrix depending on $\psi_j$ and the hyperparameters $\bmu_{\bLambda}$ and $\bSigma_{\bLambda}$, being a $p\times 1$ mean vector and a symmetric positive definite matrix of size $p\times p$. The $p\times p$ covariance matrix $\bSigma$ has an Inverse G-Wishart distribution with fully connected graph $\Gfull$, shape parameter $\xi_{\bSigma}>0$ and a $p\times p$ symmetric positive definite scale matrix $\bLambda_{\bSigma}$. The remaining hyperparamters are $\kappa_{\psi},\delta_{\psi}>0$. 

For the application studied in this section, we use
\begin{equation}
    \bLambda=\underset{k=1,\ldots,p}{\mbox{blockdiag}}(\blambda_k),\,\,\mbox{with}\,\,\blambda_k=\left[\begin{array}{c}
    \lambda_{k1}\\
    \vdots\\
    \lambda_{km_k}
    \end{array}\right],\,\, k=1,\ldots,p,
    \label{eq:LambdaDef}
\end{equation}
and set $\lambda_{k1}$ to 1 for identifiability.
Suppose, for instance, that the latent factors are $p=3$ and $m=m_1+m_2+m_3$, with $m_1=3$, $m_2=4$ and $m_3=3$. Then
\begin{equation*}
    \bLambda=\left[\begin{array}{ccc}
    1 & 0 & 0\\
    \lambda_{12} & 0 & 0\\
    \lambda_{13} & 0 & 0\\
    0 & 1 & 0\\
    0 & \lambda_{22} & 0\\
    0 & \lambda_{23} & 0\\
    0 & \lambda_{24} & 0\\
    0 & 0 & 1\\
    0 & 0 & \lambda_{32}\\
    0 & 0 & \lambda_{33}\\
    \end{array}\right]\quad\mbox{and}\quad\bdeta_i=\left[\begin{array}{c}
    \eta_{i1}\\
    \eta_{i2}\\
    \eta_{i3}
    \end{array}\right].
\end{equation*}
Given the structure of $\bLambda$ provided in \eqref{eq:LambdaDef}, we can replace the prior specification on $\bLambda$ of model \eqref{eq:complexSEM} with
\begin{equation*}
    \lambda_{kj'}\,\vert\,\psi_{kj'}\simind N(\mu_{\lambda},\sigma^2_\lambda\psi_{kj'}),\,\,\, k=1,\ldots,p,\,\,\, j'=2,\ldots,m_k,
\end{equation*}
with $\lambda_{kj'}$ and $\psi_{kj'}$ respectively denoting the elements of $\blambda$ and $\bpsi$ corresponding to the $k$th unobservable component and its associated $j'$th observable outcome, for $k=1,\ldots,p$ and $j'=1,\ldots,m_k$. The hyperparameters $\mu_{\lambda}$ and $\sigma^2_\lambda>0$ are user-specified.

Similarly to the notation defined for the entries of $\blambda$ and $\bpsi$, let $\nu_{kj'}$ denote a single intercept and $\bdeta=[\bdeta_1,\ldots,\bdeta_n]$.
In order to derive a tractable MFVB approximation for fitting model \eqref{eq:simpleSEM}, we impose the approximating density factorization
\begin{align}
\begin{split}
    q(\bnu,\bLambda,\bdeta,\bpsi,\bSigma)&=q(\bnu)q(\blambda)q(\bpsi)q(\bSigma)\prod_{i=1}^nq(\bdeta_i)\\
    &=q(\bSigma)\prod_{k=1}^p\prod_{j'=1}^{m_k}\{q(\nu_{kj'})q(\psi_{kj'})\}\prod_{k=1}^p\prod_{j'=2}^{m_k} q(\lambda_{kj'})\prod_{i=1}^nq(\bdeta_i).
\end{split}
    \label{eq:compSEMrestr}
\end{align}
The optimal approximating densities arising from \eqref{eq:compSEMrestr} are then:
\begin{equation*}
    \begin{array}{c}
    q^*(\nu_{kj'})\,\,\,\mbox{is}\,\,\,N\big(\mu_{q(\nu_{kj'})},\sigma^2_{q(\nu_{kj'})}\big),\,\,\,k=1,\ldots,p,\,\,j'=1,\ldots,m_k,\\[1ex]
    \mbox{with}\,\,\,\mu_{q(\nu_{kj'})}\equiv \sigma^2_{q(\nu_{kj'})}\mu_{q(1/\psi_{kj'})}\sum_{i=1}^n\big((\by_i)_{kj'}-\mu_{q(\lambda_{kj'})}\mu_{q(\eta_{ik})}\big)\,\,\,\mbox{and}\\[1ex]
    \sigma^2_{q(\nu_{kj'})}\equiv1\big/(n\mu_{q(1/\psi_{kj'})}+1/\sigma^2_\nu);
    \end{array}
\end{equation*}
\begin{equation*}
    \begin{array}{c}
    q^*(\lambda_{kj'})\,\,\,\mbox{is}\,\,\,N\big(\mu_{q(\lambda_{kj'})},\sigma^2_{q(\lambda_{kj'})}\big),\,\,\,k=1,\ldots,p\,\,j'=2,\ldots,m_k,\\[1ex]
    \mbox{with}\,\,\,\mu_{q(\lambda_{kj'})}\equiv\sigma^2_{q(\lambda_{kj'})}\left[\sum_{i=1}^n\big\{\mu_{q(\eta_{ik})}\big((\by_i)_{kj'}-\mu_{q(\nu_{kj'})}\big)\big\}+\mu_\lambda/\sigma^2_\lambda\right]\mu_{q(1/\psi_{kj'})}\,\,\,
    \mbox{and}\\[1ex]
    \sigma^2_{q(\lambda_{kj'})}\equiv1\big/\big\{\mu_{q(1/\psi_{kj'})}\big(\sum_{i=1}^n\mu_{q(\eta_{ik}^2)}+1/\sigma^2_{\lambda}\big)\big\};
    \end{array}
\end{equation*}
\begin{equation*}
    \begin{array}{c}
    q^*(\bdeta_i)\,\,\,\mbox{is}\,\,\,N\big(\bmu_{q(\bdeta_i)},\bSigma_{q(\bdeta_i)}\big),\,\,\,i=1,\ldots,n,\\[1ex]
    \mbox{with}\,\,\,\bmu_{q(\bdeta_i)}\equiv\bSigma_{q(\bdeta_i)}\left[\begin{array}{c}
    \sum_{j'=1}^{m_1}\mu_{q(\lambda_{1j'})}\mu_{q(1/\psi_{1j'})}\big((\by_i)_{1j'}-\mu_{q(\nu_{1j'})}\big) \\
    \vdots \\ \sum_{j'=1}^{m_p}\mu_{q(\lambda_{pj'})}\mu_{q(1/\psi_{pj'})}\big((\by_i)_{pj'}-\mu_{q(\nu_{pj'})}\big)
    \end{array}\right]\,\,\,\mbox{and}
    \\[1ex]
    \bSigma_{q(\bdeta_i)}\equiv\left(\diag\left(\left[\begin{array}{ccc}
    \sum_{j'=1}^{m_1}\mu_{q(\lambda_{1j'}^2)}\mu_{q(1/\psi_{1j'})} & \ldots & \sum_{j'=1}^{m_p}\mu_{q(\lambda_{pj'}^2)}\mu_{q(1/\psi_{pj'})}
    \end{array}\right]^T\right)+\bM_{q(\bSigma^{-1})}\right)^{-1};\\[2ex]
    \end{array}
\end{equation*}
\begin{equation*}
    \begin{array}{c}
    q^*(\psi_{kj'})\,\,\,\mbox{is}\,\,\,\mbox{Inverse-}\chi^2\big(\kappa_{q(\psi_{kj'})},\delta_{q(\psi_{kj'})}),\,\,\,k=1,\ldots,p,\,\,j'=1,\ldots,m_k,\\[1ex]
    \mbox{with}\,\,\,\kappa_{q(\psi_{kj'})}\equiv  n+\kappa_\psi+1\,\,\,
    \mbox{and}\\[1ex]\delta_{q(\psi_{kj'})}\equiv\sum_{i=1}^n\big((\by_i)_{kj'}^2+\mu_{q(\nu_{kj'}^2)}+\mu_{q(\lambda_{kj'}^2)}\mu_{q(\eta_{ik}^2)}-2(\by_i)_{kj'}\mu_{q(\nu_{kj'})}-2(\by_i)_{jk'}\mu_{q(\lambda_{kj'})}\mu_{q(\eta_{ik})}\\[1ex]
    +2\mu_{q(\nu_{kj'})}\mu_{q(\lambda_{kj'})}\mu_{q(\eta_{ik})}\big)+\frac{1}{\sigma^2_\lambda}\big(\mu_{q(\lambda_{kj'}^2)}-2\mu_{\lambda}\mu_{q(\lambda_{kj'})}+\mu^2_\lambda\big)+\delta_{\psi};\\[2ex]
    \end{array}
\end{equation*}
\begin{equation*}
    \begin{array}{c}
    q^*(\bSigma)\,\,\,\mbox{is}\,\,\,\mbox{Inverse G-Wishart}\big(\Gfull,\xi_{q(\bSigma)},\bLambda_{q(\bSigma)}\big),\\[1ex]
    \mbox{with}\,\,\,\xi_{q(\bSigma)}\equiv n+\xi_{\bSigma}\,\,\,\mbox{and}\,\,\,\bLambda_{q(\bSigma)}\equiv\sum_{i=1}^n\big(\bSigma_{q(\bdeta_i)}+\bmu_{q(\bdeta_i)}\bmu_{q(\bdeta_i)}^T\big)+\bLambda_{\bSigma}.
    \end{array}
\end{equation*}
Derivational details about these approximating densities are provided in the appendix. 

The MFVB scheme for fitting model \eqref{eq:complexSEM} under the density product restriction \eqref{eq:compSEMrestr} is presented as Algorithm \ref{alg:MFVBcompSEM}. The expressions for $\mu_{q(\nu_{kj'}^2)}$, $\mu_{q(\lambda_{kj'}^2)}$, $\mu_{q(1/\psi_{kj'})}$, $k=1,\ldots,p$ and $j'=1,\ldots,m_k$, $\bmu_{q(\bdeta_i^2)}=[\mu_{q(\eta_{i1}^2)},\ldots,\mu_{q(\eta_{ip}^2)}]$, $i=1,\ldots,n$, and $\bM_{q(\bSigma^{-1})}$ are given as updates of this algorithm.

\begin{algorithm}[!ht]
	\begin{center}
		\begin{minipage}[t]{154mm}
			\begin{small}
				\begin{itemize}
					\setlength\itemsep{4pt}
					\item[] \textbf{Data Input:} $\by_i=[(\by_i)_{11},\ldots,(\by_i)_{1m_1},\ldots,(\by_i)_{p1},\ldots,(\by_i)_{pm_p}]$, $i=1,\ldots,n$, vectors of length $m=\sum_{k=1}^p m_k$.
					\item[] \textbf{Hyperparameter Input:} $\mu_{\lambda}\in\mathbb{R}$, $\sigma_\nu,\sigma_\lambda,\kappa_\psi,\delta_\psi\in\mathbb{R}^+$, $\xi_{\bSigma}>2p-2$ and $\bLambda_{\bSigma}$ a $p\times p$ symmetric positive definite matrix.
					\item[] \textbf{Initialize:} $\mu_{q(\nu_{kj'})}$, $\mu_{q(\nu_{kj'}^2)},\mu_{q(1/\psi_{kj'})}\in\mathbb{R}^+$, $k=1,\dots,p$, $j'=1,\ldots,m_k$; $\mu_{q(\lambda_{kj'})}$, $\mu_{q(\lambda_{kj'}^2)}\in\mathbb{R}^+$, $k=1,\dots,p$, $j'=2,\ldots,m_k$; $\bmu_{q(\bdeta_i)}$, $\bmu_{q(\bdeta_i^2)}$ vectors of length $p$ with elements $\in\mathbb{R}^+$ $i=1,\ldots,n$; $\bM_{q(\bSigma^{-1})}$ a $p\times p$ symmetric positive definite matrix.
					\item[] \textbf{Set:}  $\mu_{q(\lambda_{k1})}=\mu_{q(\lambda_{k1}^2)}=1$, $k=1,\ldots,p$; $\xi_{q(\bSigma)}= n+\xi_{\bSigma}$; $\kappa_{q(\psi_{kj'})} = n+\kappa_\psi+1$, $k=1,\ldots,p$, $j' = 1,\dots,m_k$.
					\item[] \textbf{Cycle until convergence:}
					\begin{itemize}
						\setlength\itemsep{4pt}
						\item[] For $k=1,\ldots,p$ and $j'=1,\ldots,m_k$:
						\begin{itemize}
						    \setlength\itemsep{4pt}
    					    \item[] $\sigma^2_{q(\nu_{kj'})}\longleftarrow1\big/(n\mu_{q(1/\psi_{kj'})}+1/\sigma^2_\nu)$
    					    \item[] $\mu_{q(\nu_{kj'})}\longleftarrow \sigma^2_{q(\nu_{kj'})}\mu_{q(1/\psi_{kj'})}\sum_{i=1}^n\big((\by_i)_{kj'}-\mu_{q(\lambda_{kj'})}\mu_{q(\eta_{ik})}\big)$
    					    \item[] $\mu_{q(\nu_{kj'}^2)}\longleftarrow\sigma^2_{q(\nu_{kj'})}+\mu_{q(\nu_{kj'})}^2$
    					    \item[] $\delta_{q(\psi_{kj'})}\longleftarrow\sum_{i=1}^n\big((\by_i)_{kj'}^2+\mu_{q(\nu_{kj'}^2)}+\mu_{q(\lambda_{kj'}^2)}\mu_{q(\eta_{ik}^2)}-2(\by_i)_{kj'}\mu_{q(\nu_{kj'})}$
                            \item[]$\qquad\qquad\qquad-2(\by_i)_{jk'}\mu_{q(\lambda_{kj'})}\mu_{q(\eta_{ik})}+2\mu_{q(\nu_{kj'})}\mu_{q(\lambda_{kj'})}\mu_{q(\eta_{ik})}\big)$
                            \item[]$\qquad\qquad\quad+\frac{1}{\sigma^2_\lambda}\big(\mu_{q(\lambda_{kj'}^2)}-2\mu_{\lambda}\mu_{q(\lambda_{kj'})}+\mu^2_\lambda\big)+\delta_{\psi}$
    					    \item[] $\mu_{q(1/\psi_{kj'})}\longleftarrow \kappa_{q(\psi_{kj'})}/\delta_{q(\psi_{kj'})}$
    					   	\item[] If $j'>1$:
						       \begin{itemize}
						        \setlength\itemsep{4pt}
    					            \item[]  $\sigma^2_{q(\lambda_{kj'})}\longleftarrow1\big/\big\{\mu_{q(1/\psi_{kj'})}\big(\sum_{i=1}^n\mu_{q(\eta_{ik}^2)}+1/\sigma^2_{\lambda}\big)\big\}$
    					            \item[] $\mu_{q(\lambda_{kj'})}\longleftarrow\sigma^2_{q(\lambda_{kj'})}\left[\sum_{i=1}^n\big\{\mu_{q(\eta_{ik})}\big((\by_i)_{kj'}-\mu_{q(\nu_{kj'})}\big)\big\}+\mu_\lambda/\sigma^2_\lambda\right]\mu_{q(1/\psi_{kj'})}$
    					            \item[] $\mu_{q(\lambda_{kj'}^2)}\longleftarrow\sigma^2_{q(\lambda_{kj'})}+\mu_{q(\lambda_{kj'})}^2$
    					       \end{itemize}
    			        \end{itemize}
						\item[] For $i=1,\ldots,n$:
						\begin{itemize}
						    \setlength\itemsep{4pt}
    					    \item[] $\bSigma_{q(\bdeta_i)}\longleftarrow\left(\diag\left(\left[\begin{array}{c}
                            \sum_{j'=1}^{m_1}\mu_{q(\lambda_{1j'}^2)}\mu_{q(1/\psi_{1j'})}\\
                            \vdots\\
                            \sum_{j'=1}^{m_p}\mu_{q(\lambda_{pj'}^2)}\mu_{q(1/\psi_{pj'})}
                            \end{array}\right]\right)+\bM_{q(\bSigma^{-1})}\right)^{-1}$
                            \item[] $\bmu_{q(\bdeta_i)}\longleftarrow\bSigma_{q(\bdeta_i)}\left[\begin{array}{c}
                              \sum_{j'=1}^{m_1}\mu_{q(\lambda_{1j'})}\mu_{q(1/\psi_{1j'})}\big((\by_i)_{1j'}-\mu_{q(\nu_{1j'})}\big)\\
                              \vdots\\ \sum_{j'=1}^{m_p}\mu_{q(\lambda_{pj'})}\mu_{q(1/\psi_{pj'})}\big((\by_i)_{pj'}-\mu_{q(\nu_{pj'})}\big)
                           \end{array}\right]$
                            \item[] $\bmu_{q(\bdeta_i^2)}\longleftarrow\diag\left(\bSigma_{q(\bdeta_i)}+\bmu_{q(\bdeta_i)}\bmu_{q(\bdeta_i)}^T\right)$
    			        \end{itemize}
    			        \item[] $\bLambda_{q(\bSigma)}\longleftarrow\sum_{i=1}^n\big(\bSigma_{q(\bdeta_i)}+\bmu_{q(\bdeta_i)}\bmu_{q(\bdeta_i)}^T\big)+\bLambda_{\bSigma}$
    			        \item[] $\bM_{q(\bSigma^{-1})}\longleftarrow(\xi_{q(\bSigma)}-p+1)\bLambda_{q(\bSigma)}^{-1}$
					\end{itemize}
					\item[] \textbf{Output:} $\mu_{q(\nu_{kj'})}$, $\sigma^2_{q(\nu_{kj'})}$, $\mu_{q(\lambda_{kj'})}$, $\sigma^2_{q(\lambda_{kj'})}$, $\kappa_{q(\psi_{kj'})}$, $\delta_{q(\psi_{kj'})}$, $k=1,\ldots,p$, $j'=1,\ldots,m_k$; $\bmu_{q(\bdeta_i)}$, $\bSigma_{q(\bdeta_i)}$, $i=1,\ldots,n$; $\xi_{\bSigma}$, $\bLambda_{\bSigma}$.
				\end{itemize}
			\end{small}
		\end{minipage}
	\end{center}
	\caption{\textit{Algorithm for fitting model \eqref{eq:complexSEM} via MFVB, under restriction \eqref{eq:compSEMrestr}.}}
	\label{alg:MFVBcompSEM}
\end{algorithm}

\begin{figure}[!ht]
	\centering
	{\includegraphics[width=0.9\textwidth]{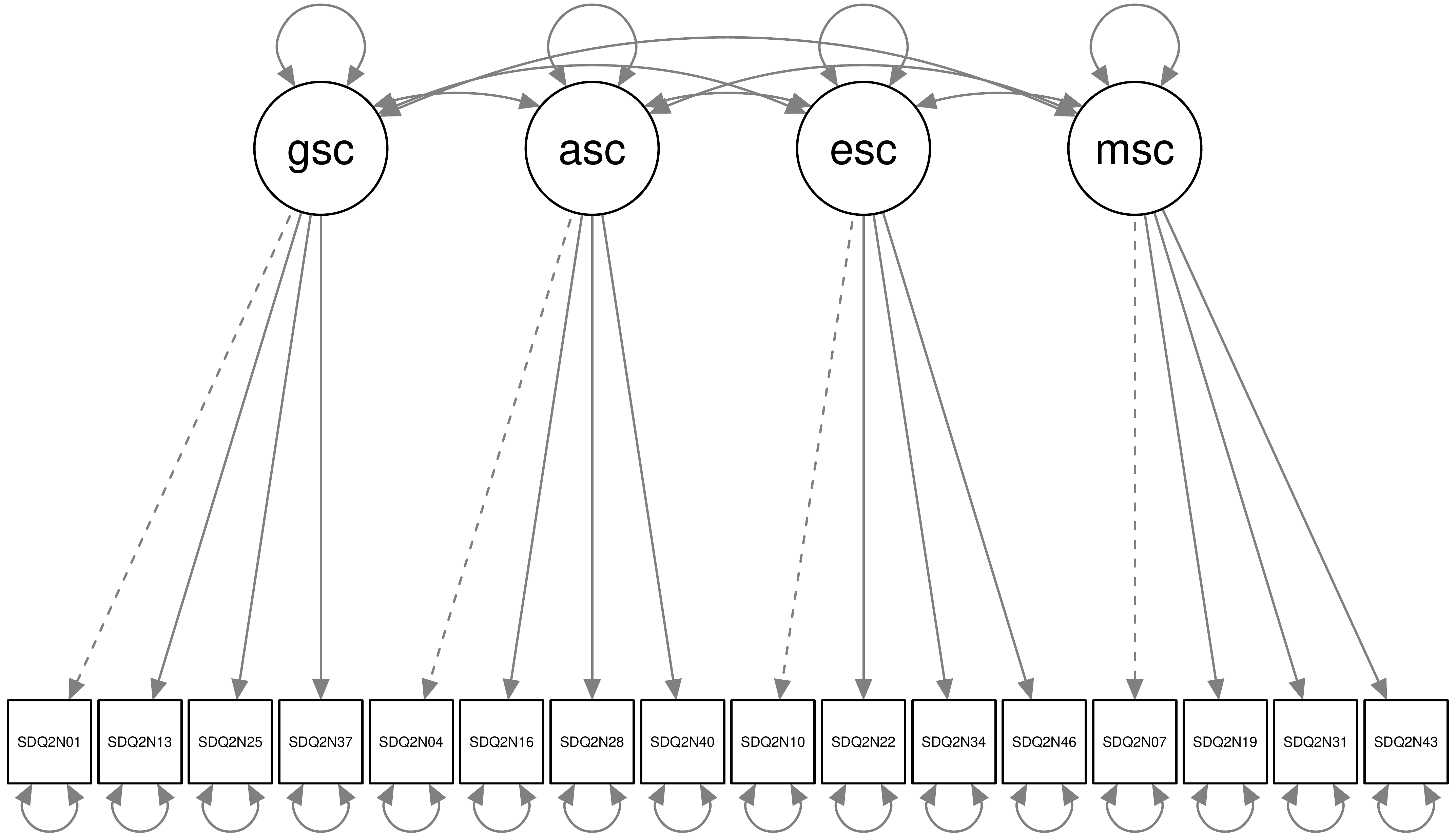}}
	\caption{\it Path diagram of model \eqref{eq:complexSEM} for the  self-concept data examined in \cite{byrne2016structural}. The observed outcomes are shown in the squares. The unobservable factors represented by circles are general self-concept (GSC), academic self-concept (ASC), English self-concept (ESC) and mathematics self-concept (MSC). Circular arrows represent correlations, whereas the dashed lines indicate the corresponding latent factor loadings are set to 1 for identifiability.}
	\label{fig:complexSemPathDiagram} 
\end{figure}

We employed Algorithm \ref{alg:MFVBcompSEM} and fitted model \eqref{eq:complexSEM} to the self-concept data studied in \cite{byrne2016structural}. The dataset was collected from 265 early adolescents in grade 7 and consists of 16 observed variables from four subscales of the Self Description Questionnaire II \citep{byrne2016structural,marsh1992self}.
A confirmatory factor analysis model was used to test the hypothesis that self-concept (SC) is a multidimensional construct composed of $p=4$ intercorrelated factors: general SC (GSC), academic SC (ASC), English SC (ESC), and mathematics SC (MSC). The corresponding path diagram is displayed as Figure \ref{fig:complexSemPathDiagram}, where the 16 observed variables ($m_1=m_2=m_3=m_4=4$) identified by codes such as SDQ2N01 and SDQ2N43 are represented by rectangles.

\begin{figure}[!ht]
	\centering
	{\includegraphics[width=1\textwidth]{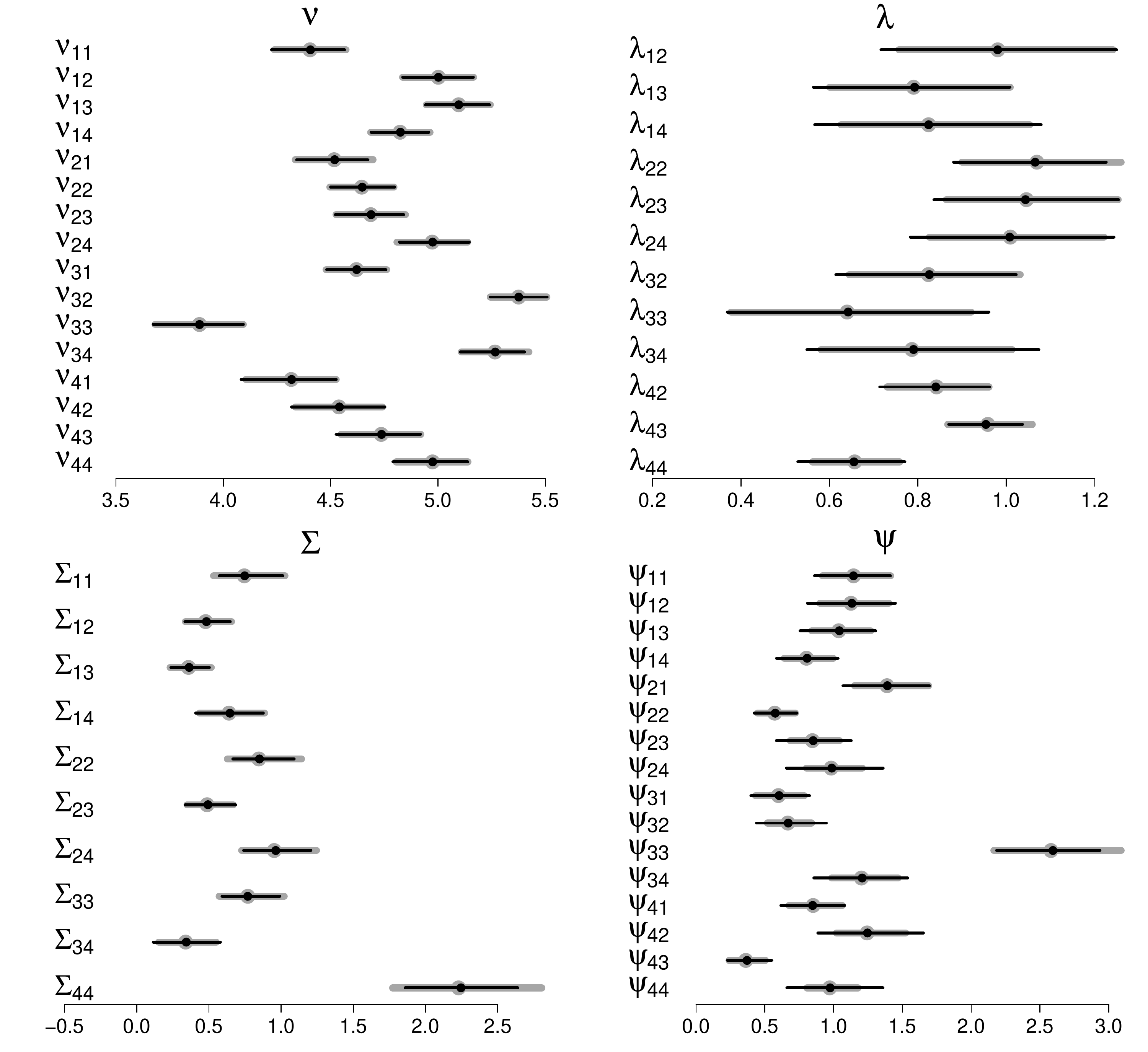}}
	\caption{\it Visualization of 95\% credible intervals for the parameters of model \eqref{eq:complexSEM} applied to the self-concept data. The black lines correspond to percentile bootstrap credible intervals obtained via MFVB and 1,000 bootstrap iterations. These are compared to the MCMC benchmark provided by thicker gray lines.}
	\label{fig:SCmultifact} 
\end{figure}
We ran MFVB via Algorithm \ref{alg:MFVBcompSEM} and MCMC in \textsf{rstan} using $\mu_{\lambda}= 0$, $\sigma_\nu = 10,\sigma_\lambda =1,\kappa_\psi =1,\delta_\psi = 0.01$, $\xi_{\bSigma}=2p+10$ and $\bLambda_{\bSigma} = 10\bI$. For MCMC, we generated 3 chains, each of length 20,000 and with burn-in of length 10,000. The credible intervals for each parameter of interest obtained from both methods are shown in Figure \ref{fig:SCmultifact}. The MFVB credible intervals were produced using 1,000 bootstrap iterations and the percentile approach. These results demonstrate that MFVB supported by bootstrap achieve performances very similar to those provided by the MCMC benchmark. The MFVB point estimates do not show substantial bias and the MFVB intervals present coverage very similar to that of their MCMC equivalent. For some of the $\blambda$ entries, the MFVB credible intervals are slightly wider than the MCMC target.

\section{Conclusion}\label{sec:conclusion}

This paper studies a mean field variational Bayes approach for fitting confirmatory factor analysis models with one or more latent factors. We have shown that this variational inference method is extremely fast and accurate, when paired with a suitable resampling strategy, and offers a particularly attractive alternative to standard Markov chain Monte Carlo.

To deal with the typical lack of accuracy of MFVB posterior density approximations, we have studied the use of bootstrap and shown improved variational inference for the model parameters of interest. Our results based on real and simulated examples indicate that the coverage of intervals produced with the bootstrap corrections is very close to the advertised level. Also simple jackknife resampling was shown to greatly improve the empirical coverage of credible intervals directly derived from MFVB approximating densities. Pivotal credible intervals may be second order accurate if a consistent estimator of the variance of the variational inference estimator is used. When this is not available, a consistent estimator can be obtained via iterated bootstrap, that is by bootstrapping each bootstrap dataset to compute the standard error of the variational inference estimator obtained from each dataset bootstrapped at the first level. However, this strategy is associated with much higher computational costs, therefore it has not been explored in our examples.

The variational approximation framework presented in this paper can be extended to more challenging situations. For instance, Algorithm \ref{alg:MFVBcompSEM} can be modified to treat models involving exogenous covariates or cross-loadings, or to work with incomplete data. However, we leave this for future research.

\ifthenelse{\boolean{UnBlinded}}{
\section*{Acknowledgments}
The authors were supported by the Australian Research Council Center of Excellence for Mathematical and Statistical Frontiers (CE140100049). Luca Maestrini was also supported by the Australian Research Council Discovery Project DP180100597.}


\bibliographystyle{apalike}
\bibliography{ref}

\null\vfill\eject

%
%
\appendix
\renewcommand{\thealgorithm}{A.\arabic{algorithm}}
\setcounter{algorithm}{0}
\renewcommand{\thesection}{A.\arabic{section}}
\setcounter{section}{0}

\centerline{\Large Appendix for:}
\vskip5mm
\centerline{\Large\bf Fitting Structural Equation Models via Variational Approximations}
\vskip5mm
\ifthenelse{\boolean{UnBlinded}}{
\centerline{\normalsize\sc By Khue-Dung Dang$\null^{1}$ and Luca Maestrini$\null^{2,3}$}
\vskip5mm
\centerline{\textit{$\null^1$University of Melbourne, $\null^2$University of Technology Sydney and}}
\vskip1mm
\centerline{\textit{$\null^3$Australian Research Council Centre of Excellence for Mathematical and Statistical Frontiers}}{}}

\section{Derivation of Algorithm \ref{alg:MFVBcompSEM}}

Let $\by=[\by_1,\ldots,\by_n]$. Provided the definition of $\bLambda$ in \eqref{eq:LambdaDef}, the likelihood function arising from model \eqref{eq:complexSEM} is
\begin{align*}
    &\hspace{-1cm}p(\by;\bnu,\bLambda,\bdeta,\bpsi,\bSigma)=\\
    &\prod_{i=1}^n\left[ (2\pi)^{-m/2}\prod_{k=1}^p\prod_{j'=1}^{m_k}\psi_{kj'}^{-1/2}\exp\left\{-\frac{1}{2}(\by_i-\bnu-\bLambda\bdeta_i)^T\diag(1/\bpsi)(\by_i-\bnu-\bLambda\bdeta_i)\right\}\right]\\
    \times&\prod_{k=1}^p\prod_{j'=1}^{m_k}\left\{(2\pi\sigma^2_\nu)^{-1/2}\exp\left(-\frac{\nu_{kj'}^2}{2\sigma^2_\nu}\right)\right\}\prod_{k=1}^p\prod_{j'=1}^{m_k}\left[(2\pi\sigma^2_\lambda\psi_{kj'})^{-1/2}\exp\left\{-\frac{(\lambda_{kj'}-\mu_\lambda)^2}{2\sigma^2_\lambda\psi_{kj'}}\right\}\right]\\
    \times&\prod_{k=1}^p\prod_{j'=1}^m\left[\left(\frac{\delta_\psi}{2}\right)^{\kappa_\psi/2}\left\{\Gamma\left(\frac{\kappa_\psi}{2}\right)\right\}^{-1}\psi_{kj'}^{-(\kappa_\psi+2)/2}\exp\left(-\frac{\delta_{\psi}}{2\psi_{kj'}}\right)\right]\\
    \times&\prod_{i=1}^n\left\{(2\pi)^{-p/2}|\bSigma|^{-1/2}\exp\left(-\frac{1}{2}\bdeta_i^T\bSigma^{-1}\bdeta_i\right)\right\}\\
    \times&\displaystyle{\frac{|\bLambda_{\bSigma}|^{(\xi_{\bSigma}-p+1)/2}}{2^{d(\xi_{\bSigma}-p+1)/2}\pi^{p(p-1)/4}\prod_{k=1}^p\Gamma(\frac{\xi_{\bSigma}-p-k}{2}+1)}}\,
	|\bSigma|^{-(\xi_{\bSigma}+2)/2}\exp\left\{-\frac{1}{2}\mbox{\rm tr}(\bLambda_{\bSigma}\bSigma^{-1})\right\}.
\end{align*}
Consequently, the log-likelihood function is
\begin{align*}
    &\log p(\by;\bnu,\bLambda,\bdeta,\bpsi,\bSigma)=-\frac{1}{2}\sum_{i=1}^n(\by_i-\bnu-\bLambda\bdeta_i)^T\diag(1/\bpsi)(\by_i-\bnu-\bLambda\bdeta_i)-\frac{1}{2\sigma^2_\nu}\sum_{k=1}^p\sum_{j'=1}^{m_k}\nu_{kj'}^2\\
    &-\frac{n+\xi_{\bSigma}+2}{2}\log(|\bSigma|)-\frac{1}{2}\sum_{i=1}^n\bdeta_i^T\bSigma^{-1}\bdeta-\frac{1}{2}\mbox{\rm tr}(\bLambda_{\bSigma}\bSigma^{-1})-\frac{n+\kappa_\psi+3}{2}\sum_{k=1}^p\sum_{j'=1}^{m_k}\log(\psi_{kj'})\\&
    -\frac{\delta_{\psi}}{2}\sum_{k=1}^p\sum_{j'=1}^{m_k}\frac{1}{\psi_{kj'}}-\frac{1}{2\sigma^2_\lambda}\sum_{k=1}^p\sum_{j'=1}^{m_k}\frac{(\lambda_{kj'}-\mu_\lambda)^2}{\psi_{kj'}}+\mbox{const},
\end{align*}
where `const' denotes terms not depending on the variables of interest. 

The full conditional density functions can be obtained from the log-likelihood function.
For $k=1,\ldots,p$ and $j'=1,\ldots,m_k$,
\begin{align*}
    p(\nu_{kj'}\,\vert\,\mbox{rest})&\propto\exp \left[-\frac{1}{2}\left\{\nu_{kj'}^2\left(\frac{n}{\psi_{kj'}}+\frac{1}{\sigma^2_\nu}\right)-2\frac{\nu_{kj'}}{\psi_{kj'}}\sum_{i=1}^n\left((\by_i)_{kj'}-\lambda_{kj'}\eta_{ik}\right)\right\}\right]\quad\mbox{and}\\
    p(\psi_{kj'}\,\vert\,\mbox{rest})&\propto\psi_{kj'}^{-(n+\kappa_\psi+3)/2}\exp\left[-\frac{1}{2\psi_{kj'}}\left\{\sum_{i=1}^n\left((\by_i)_{kj'}-\nu_{kj'}-\lambda_{kj'}\eta_{ik}\right)^2+\frac{(\lambda_{kj'}-\mu_\lambda)^2}{\sigma^2_\lambda}+\delta_{\psi}\right\}\right].
\end{align*}
Recalling that $\lambda_{k1}=1$, for $k=1,\ldots,p$, we have that, for $k=1,\ldots,p$ and $j'=2,\ldots,m_k$,
\begin{equation*}
 p(\lambda_{kj'}\,\vert\,\mbox{rest})\propto\exp\left(-\frac{1}{2}\left[\frac{\lambda_{kj'}^2}{\psi_{kj'}}\left(\sum_{i=1}^n\eta_{ik}^2+\frac{1}{\sigma^2_\lambda}\right)-2\frac{\lambda_{kj'}}{\psi_{kj'}}\left\{\sum_{i=1}^n \big((\by_i)_{kj'}- \nu_{kj'}\big)\eta_{ik}+\frac{\mu_\lambda}{\sigma^2_\lambda}\right\}\right]\right).
\end{equation*}
For $i=1,\ldots,n$,
\begin{equation*}
    p(\bdeta_i\,\vert\,\mbox{rest})\propto\exp\left[-\frac{1}{2}\left\{\bdeta_i^T\left(\bLambda^T\diag(1/\bpsi)\bLambda+\bSigma^{-1}\right)\bdeta_i-2\bdeta_i^T\bLambda^T\diag(1/\bpsi)(\by_i-\bnu)\right\}\right].
\end{equation*}
Notice that
\begin{equation*}
    \bLambda^T\diag(1/\bpsi)\bLambda=\diag\left(\left[\begin{array}{ccc}
    \sum_{j'=1}^{m_1}\frac{\lambda_{1j'}^2}{\psi_{1j'}} & \ldots & \sum_{j'=1}^{m_p}\frac{\lambda_{pj'}^2}{\psi_{pj'}}
    \end{array}\right]^T\right)
\end{equation*}
and that
\begin{equation*}
    \bLambda^T\diag(1/\bpsi)(\by_i-\bnu)=\left[\begin{array}{ccc}
    \sum_{j'=1}^{m_1}\frac{\lambda_{1j'}}{\psi_{1j'}}\big((\by_i)_{1j'}-\nu_{1j'}\big) & \ldots & \sum_{j'=1}^{m_p}\frac{\lambda_{pj'}}{\psi_{pj'}}\big((\by_i)_{pj'}-\nu_{pj'}\big)
    \end{array}\right]^T.
\end{equation*}
The remaining full conditional is
\begin{align*}
    p(\bSigma\,\vert\,\mbox{rest})&\propto|\bSigma|^{-(n+\xi_{\bSigma}+2)/2}\exp\left\{-\frac{1}{2}\left(\sum_{i=1}^n\bdeta_i^T\bSigma^{-1}\bdeta_i+\mbox{\rm tr}(\bLambda_{\bSigma}\bSigma^{-1})\right)\right\}.
\end{align*}
Algorithm \ref{alg:MFVBcompSEM} follows from application of (5) of \cite{ormerod2010explaining} and the proposed factorization in \eqref{eq:compSEMrestr}.

\end{document}